\documentclass[a4paper,11pt,final]{article}

\usepackage{graphicx,amsmath,amssymb,algorithm,algorithmic,textcomp,tikz-cd,float,amsfonts} % Add all your packages here
\usepackage{epsfig}

\DeclareMathAlphabet\mathbfcal{OMS}{cmsy}{b}{n}

\begin{document}

% paper title: Must keep \ \\ \LARGE\bf in it to leave enough margin.
\title{\ \\ \LARGE\bf Evolution of Cooperation for Multiple Mutant Configurations on All Regular Graphs with $N \leq 14$ players}

\author{Hendrik Richter \\
HTWK Leipzig University of Applied Sciences \\ Faculty of
Engineering\\
        Postfach 301166, D--04251 Leipzig, Germany. \\ Email: 
hendrik.richter@htwk-leipzig.de }

\maketitle

\begin{abstract}

We study the emergence of cooperation in structured populations with any arrangement of cooperators and defectors on the evolutionary graph. Using structure coefficients defined for configurations describing such arrangements of any number of mutants,  we provide results for weak selection to favor cooperation over defection on any regular graph with  $N \leq 14$ vertices. Furthermore, the properties of graphs that particularly promote cooperation are analyzed. It is shown that the number of graph cycles of certain length is a good predictor for the values of the structure coefficient, and thus a tendency to favor cooperation.  Another property of particularly cooperation--promoting regular graphs with a low degree is that they are structured to have blocks with clusters of mutants that are connected by cut vertices and/or hinge vertices.

\end{abstract}

\section{Introduction}

Describing conditions for the emergence of cooperation in structured populations is a fundamental problem in evolutionary game theory~\cite{broom13,new18,nowak06,wu13}. 
In structured populations the network describing which players interact with each other may be crucial for the fixation of a strategy. Recently, several attempts have been made to explore the  universe of interaction graphs in order to link graph properties to fixation. For a single cooperator  this question has been studied intensively and recently  relationships have been mapped  for a large variety of different interaction graphs connecting which strategy is favored with the fixation probabilities and the fixation times~\cite{allen17,moell19,pav18,tka19}. These results clarify for a single mutant the relationships between the graph structure, on the one hand,  and fixation probability and fixation time, on the other. The main findings are that generally
fixation probability and fixation time is  correlated such that a higher fixation probability comes with a higher fixation time. Within this general rule, it has further been shown that generalized stars
maximize fixation probability while minimizing fixation time,   while comet--kites 
minimize fixation probability while maximizing fixation time~\cite{moell19}.
Furthermore, 
if we allow self loops and weighted links, we may construct
arbitrarily strong amplifiers of selection~\cite{pav18}.
Compared with these findings, the problem of multiple cooperators (or more than one mutant) is far less studied. One approach uses configurations and structure coefficients~\cite{chen16} and has shown that cooperation is favored over defection under conditions which can be linked to spectral graphs measures and cooperator path length~\cite{rich19a,rich19b}.

This study deals with strategy selection for multiple mutants on evolutionary graphs and addresses two central questions. The first is to find out which interaction network modeled as a regular graph yields the largest structure coefficient and therefore is most suited to promote the evolution of cooperation. This is reported for all regular graphs with $N \leq 14$ vertices ($=$ players). This question is studied subject to three parameters, the number of players, coplayers and cooperators. Answering this question may inform designing interaction networks with prescribed abilities to promote or suppress cooperation.    As there are some trends over varying these three parameters, it appears possible to conjecture for beyond the considered parameters. The second question studied takes up the observation that there are differences in the values of the structure coefficients over regular interaction graphs and asks what makes some graphs different from others in terms of promoting the evolution of cooperation.  Our main interest is what these differences are from a graph--theoretical point of view. This goes along with identifying certain properties of regular cooperation--promoting graphs. The main result is that the number of graph cycles of certain length is a good predictor of a large value of the structure coefficient. Especially
for a smaller number of coplayers graphs that particularly promote cooperation have rather cycles with small length. Furthermore, these graphs are structured to have blocks that are connected by cut vertices and/or hinge vertices. Cooperators cluster on these blocks and serve as a mutant family that may invade the remaining graph. 
The study presented here uses structure coefficients, which have been derived for birth--death and death--birth processes~\cite{chen16}. However, as the structure coefficients solely depend on the distribution of cooperators and defectors on the evolutionary graph, they could be, at least in principle, also calculated for other strategy updating processes as long as these processes are not completely random. Thus, the methodology reported here is also applicable for other types of non--imitative dynamics.

The paper is structured as follows. In Sec. 2 the main results are given. In particular, upper and lower bounds on the structure coefficients are presented for all interaction networks modelled as regular graphs with $N\leq 14$ players. Furthermore, it is shown that between maximal structure coefficients (and thus conditions favoring the prevalence of cooperation) and the count of cycles with certain length, there is an approximately linear relationship.
The results are discussed in Sec. 3, while the Appendices review the methodological framework of configurations, regular graphs and structure coefficients, discuss  graph isomorphism,  and give a collection of graphs with maximal structure coefficients.  

\section{Evolution of Cooperation}

\subsection{Upper and lower bounds on the structure coefficients}

The structure coefficient $\sigma(\pi,\mathcal{G})$ introduced by Chen et al.~\cite{chen16} (see~\cite{rich19a,rich19b} for further analysis) is a measure of whether or not cooperation is favored over defection in games with any arrangement of cooperators and defectors on regular evolutionary graphs. More strictly speaking,  in an evolutionary game with weak selection and a payoff matrix (\ref{eq:payoff}), the fixation probability of cooperation is larger than the fixation probability of defection if $\sigma(\pi,\mathcal{G}) (a-d)> (c-b)$, see also Appendix 1. This condition connects the values of the payoff matrix, the structure of the evolutionary graph $\mathcal{G}$ and the arrangement of cooperators and defectors on this graph expressed by the configuration $\pi$ with long--term prevalence of cooperation. The structure coefficient $\sigma(\pi,\mathcal{G})$ generalizes the structure coefficient $\sigma$ introduced by Tarnita et al.~\cite{tarnita09} which yields the same condition for favoring cooperation,  $\sigma (a-d)> (c-b)$, but  applies to a single cooperator (or a single mutant). By contrast,  $\sigma(\pi,\mathcal{G})$  is valid for any arrangement of cooperators and defectors on the evolutionary graph and specifically for several cooperators (or multiple mutants).

As the structure coefficient  varies over configurations $\pi$ and graphs $\mathcal{G}$, it is natural to ask about upper and lower bounds of $\sigma(\pi,\mathcal{G})$. 
In this paper, we approach this question by checking all  $\sigma(\pi,\mathcal{G})$, which appears feasible for a small number of players $N \leq 14$ and all regular graphs with up to 14 vertices. 
We classify the structure coefficients and graphs  with respect to the number of players $N$.  Furthermore, the configurations $\pi$ are also grouped according to the number of cooperators $c(\pi)$, $2 \leq c(\pi) \leq N-2$, while the graphs $\mathcal{G}$ are sorted according to the number of coplayers  $k$ (which equals the degree of the graph). As the structure coefficients $\sigma(\pi,\mathcal{G})$ vary over configurations \emph{and} graphs $\mathcal{G}$, we may define two bounds. A first is over all $2^N-2$ non--absorbing configurations, which we call $\sigma_{max_i}$. Thus, we obtain for each graph   $\mathcal{G}_i$, $i=1,2,\ldots \mathcal{L}_k(N)$, the quantity $\sigma_{max_i}=\underset{\pi}{\max} \: \sigma(\pi,\mathcal{G}_i)$. A second bound, called $\sigma_{max}$, is derived from the first bound and additionally collects over all $\mathcal{L}_k(N)$ regular graphs with a given $N$ and $k$ according to Tab.  \ref{tab:graphs}. Thus, we get 
$\sigma_{max}=\underset{i}{\max} \: \sigma_{max_i}$. For the minimum, the bounds are defined like--wise.

\begin{figure}[tb]
\includegraphics[trim = 5mm 0mm 18mm 0mm,clip,width=6.25cm, height=4.5cm]{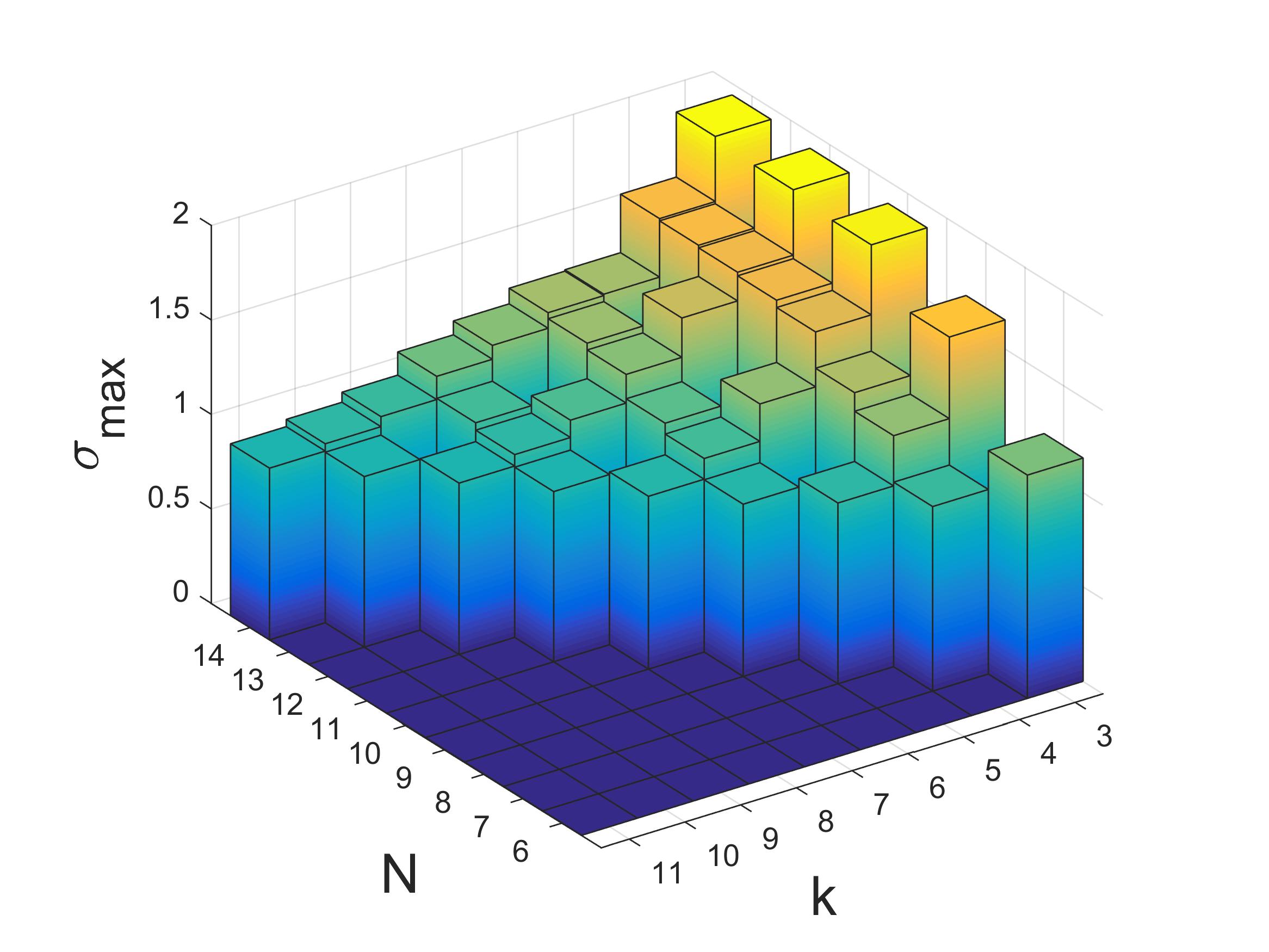}
\includegraphics[trim = 5mm 0mm 18mm 0mm,clip,width=6.25cm, height=4.5cm]{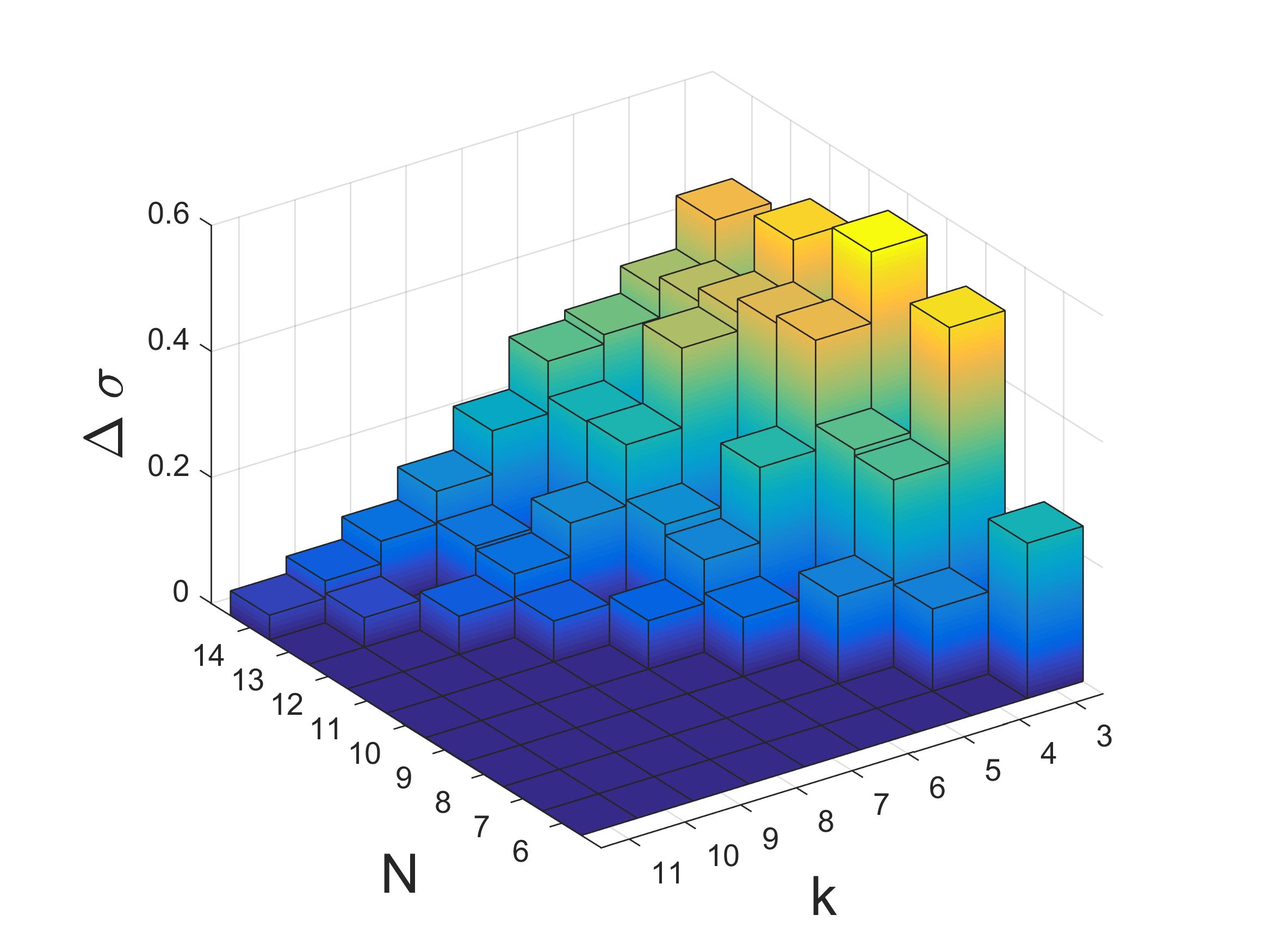}

\hspace{1cm}(a)  \hspace{5cm} (b) 

\caption{The maximal structure coefficient $\sigma_{max}$ and the maximal difference $\Delta \sigma=\sigma_{max}-\sigma_{min}$ over the number of players $N$ and coplayers $k$ for all regular interaction graphs with $6 \leq N\leq 14$ and $3 \leq k \leq N-3$ according to Tab. \ref{tab:graphs}.}
\label{fig:sigmax1}
\end{figure}

Fig. \ref{fig:sigmax1} shows the maximal structure coefficient $\sigma_{max}$ and the maximal difference $\Delta \sigma=\sigma_{max}-\sigma_{min}$ over players $N$ and coplayers $k$.   As discussed in Appendix 2 these results  apply to any instance of a regular graph, for example to random regular graphs.   
It can be seen that the maximal structure coefficient $\sigma_{max}$ is largest for $k=3$, which is cubic graphs. For $k>3$, the values of $\sigma_{max}$ get gradually smaller. In other words, the more coplayers there are, the smaller is $\sigma_{max}$.  Also,  for a constant number of coplayers, $\sigma_{max}$ increases with $N$, which is the number of players. The increase, however, gets gradually smaller and converges for $N \rightarrow \infty$ to a constant, which is $\sigma(\pi,\mathcal{G}) \rightarrow \sigma=(k+1)/(k-1)$~\cite{chen16,ohts06}. For instance, for $k=3$, the structure coefficients converge to $\sigma(\pi,\mathcal{G}) \rightarrow \sigma=2$.  
\begin{figure}[tb]

\includegraphics[trim = 5mm 0mm 18mm 0mm,clip,width=6.25cm, height=4.5cm]{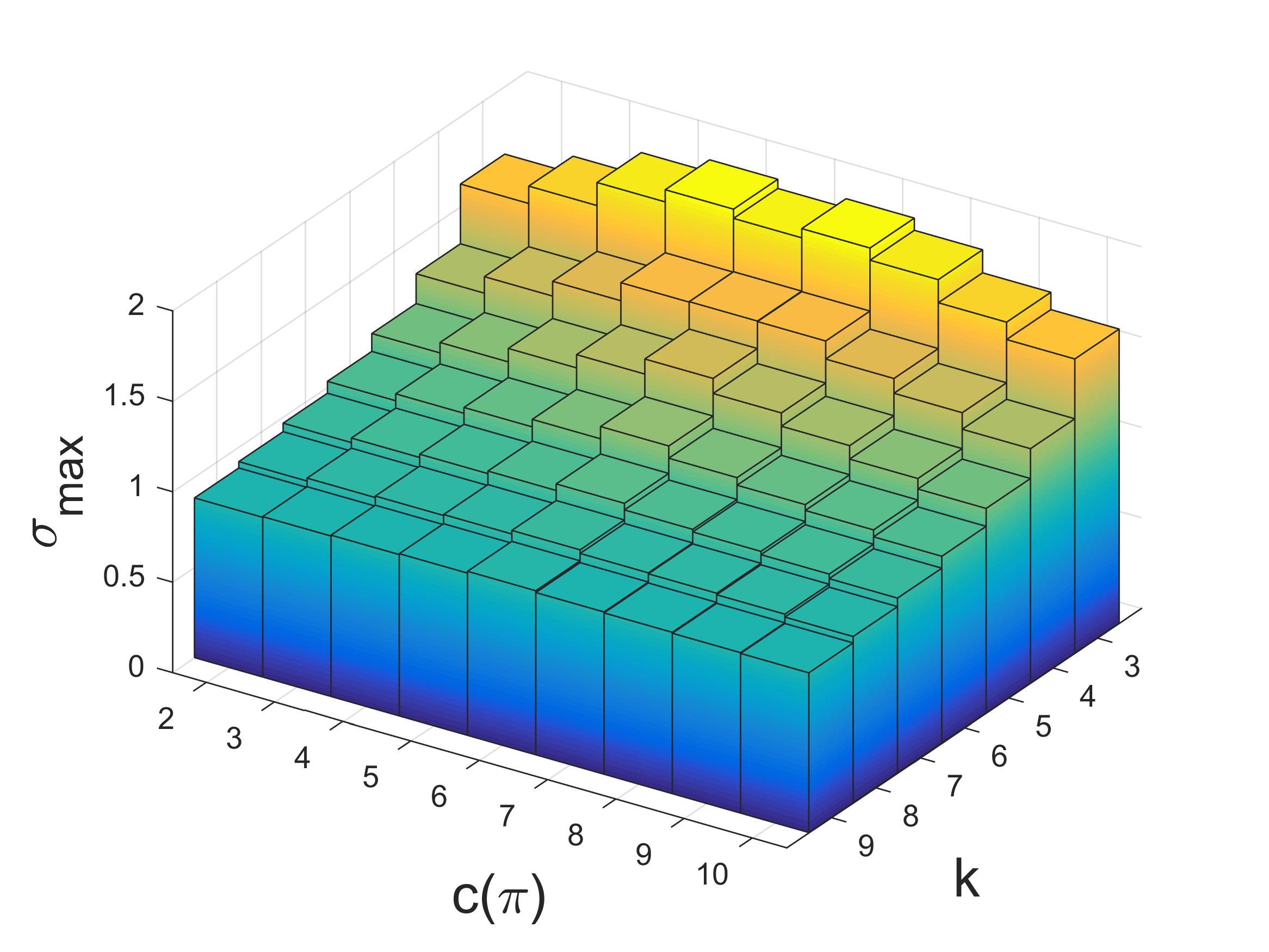}
\includegraphics[trim = 5mm 0mm 18mm 0mm,clip,width=6.25cm, height=4.5cm]{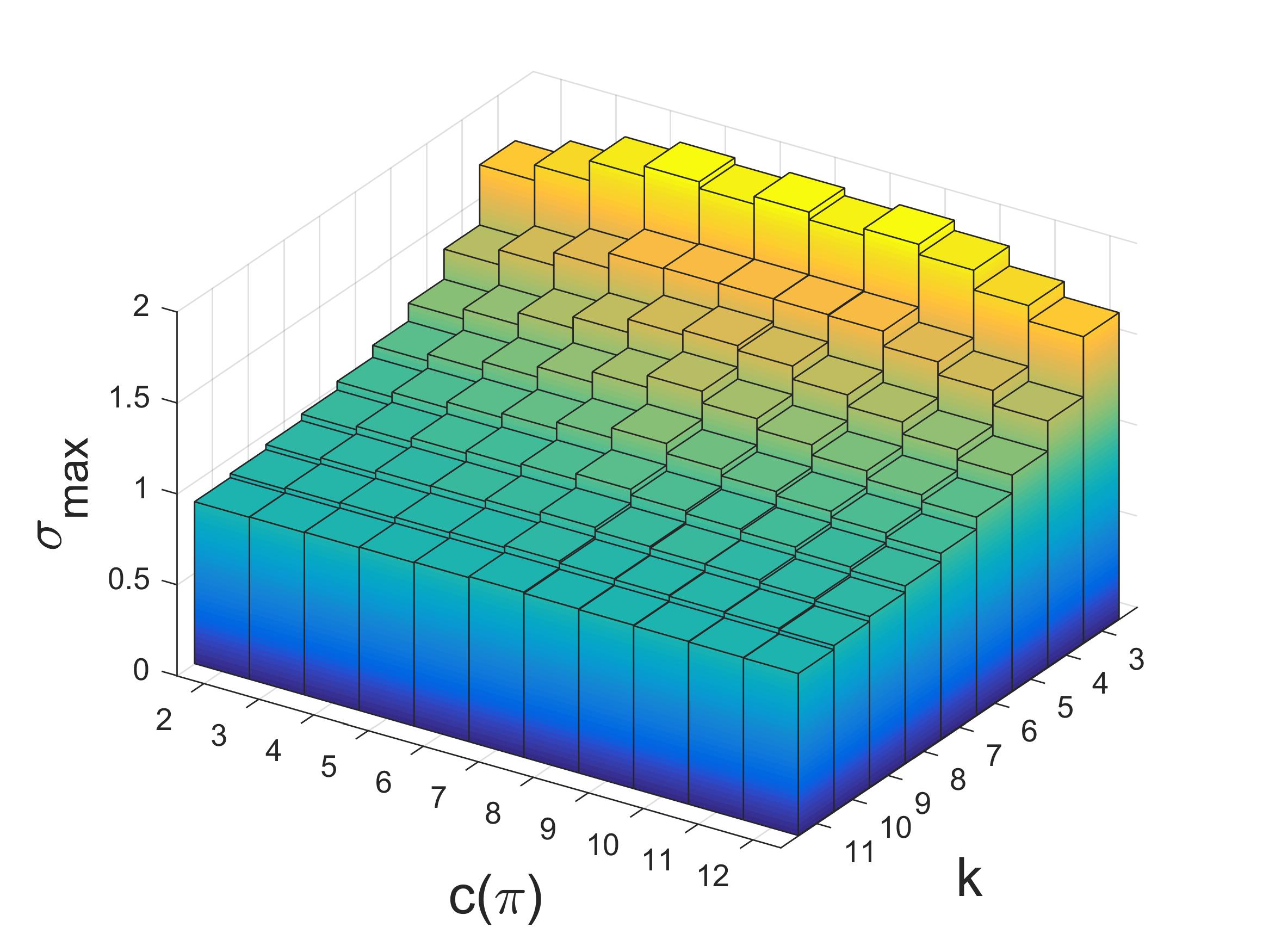}

\hspace{1cm}(a) $N=12$ \hspace{5cm} (b) $N=12$

\includegraphics[trim = 5mm 0mm 18mm 0mm,clip,width=6.25cm, height=4.5cm]{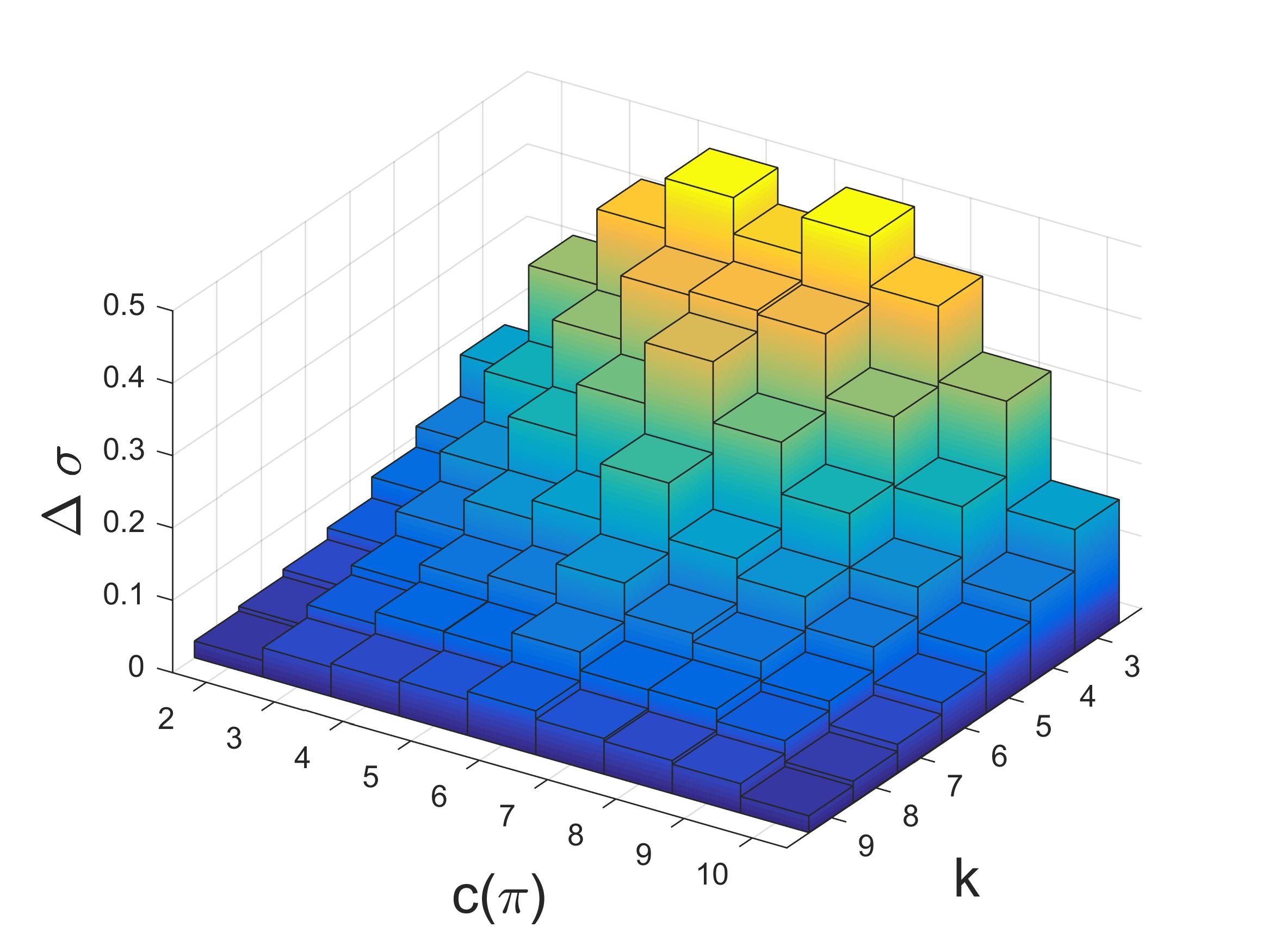}
\includegraphics[trim = 5mm 0mm 18mm 0mm,clip,width=6.25cm, height=4.5cm]{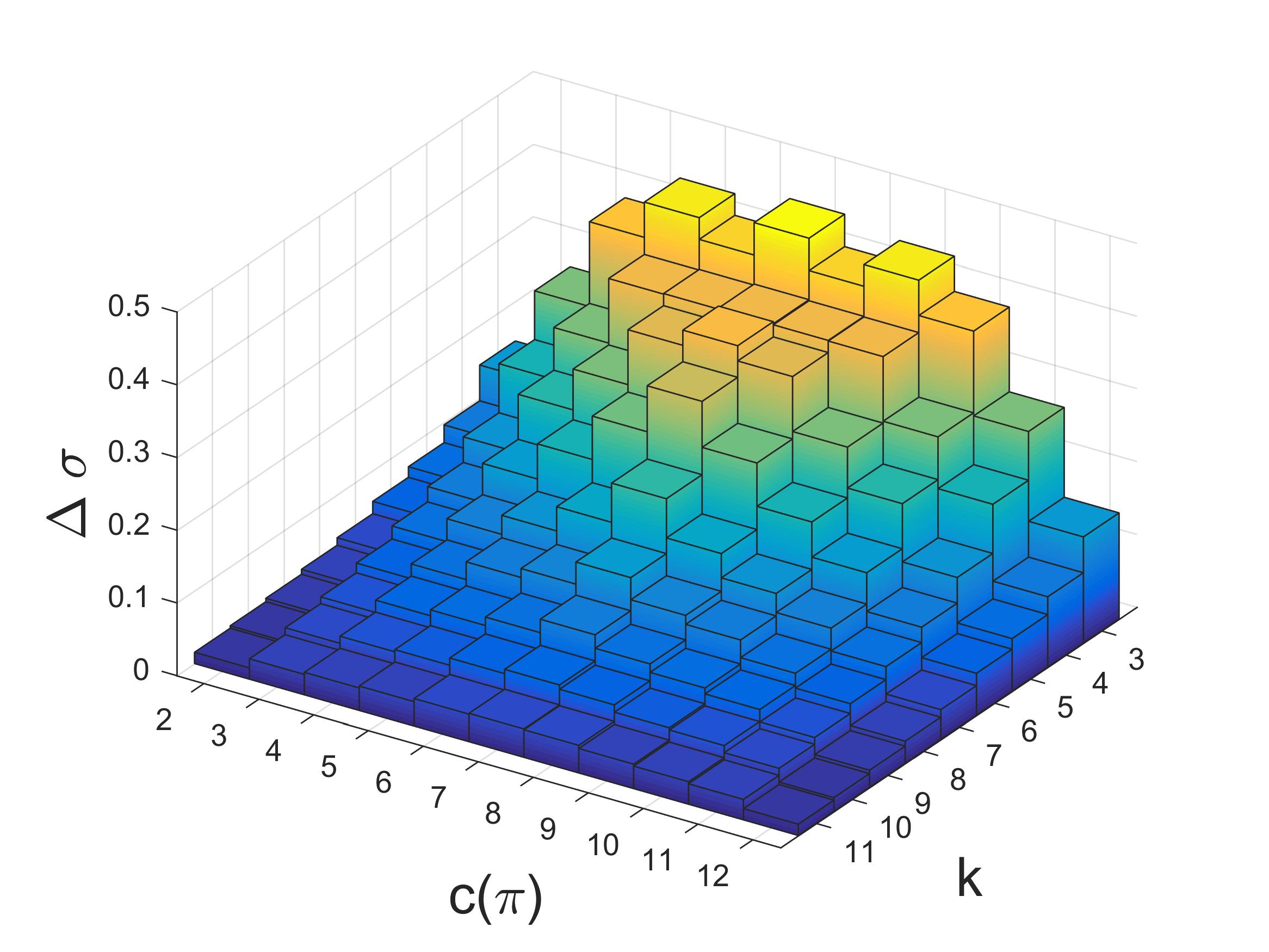}

\hspace{1cm}(c) $N=14$ \hspace{5cm} (d) $N=14$

\caption{The maximal structure coefficient $\sigma_{max}$ and the maximal difference $\Delta \sigma=\sigma_{max}-\sigma_{min}$ over the number of coplayers $k$ and cooperators $c(\pi)$ for all regular interaction graphs with $N=12$ and $N=14$ according to Tab. \ref{tab:graphs}. }
\label{fig:sigmax2}
\end{figure} 
  In other words, for the thermodynamic limit with an infinite population, prevalence of cooperation only depends on the number of coplayers $k$ of a regular graph, but not on the graph structure or the number and arrangement of cooperators on the graph. 
  The largest difference between maximal and minimal structure coefficient $\Delta \sigma=\sigma_{max}-\sigma_{min}$ we also get for $k=3$. Here, $\Delta \sigma$ increase to a largest values (for instance for $k=3$ this happens for $N=10$) before falling for $N$ getting even larger, converging to $\Delta \sigma=0$ for $N \rightarrow \infty$.  
 
We next analyze the maximal structure coefficients depending on the number of cooperators $c(\pi)$. Thus, the maximum is over all $\#_{c(\pi)}=\left(\begin{smallmatrix} N \\ c(\pi) \end{smallmatrix} \right)$ configurations with the same number of cooperators $2 \leq c(\pi) \leq N-2$
 and all regular graphs according to Tab.  \ref{tab:graphs}. The maximal values of $\sigma_{max}$ and  $\Delta \sigma$ are obtained for $c(\pi)=N/2$ for $N$ even and for both $(N+1)/2$ and $(N-1)/2$ for $N$ odd. An exception is $N=12$ and $k=3$, where $\sigma_{max}$ is obtained for $c(\pi)=5$ and $c(\pi)=7$.
Furthermore, we get the following results, see Fig. \ref{fig:sigmax2} as examples for $N=12$ and $N=14$. The value $\sigma_{max}$ and  $\Delta \sigma$ are symmetric with the number of cooperators $c(\pi)$ and generally higher for the number of cooperators and defectors exactly or approximately the same than for a small number of cooperators or a small number of defectors. For the number of coplayers $k$ getting larger, the differences over the number of cooperators $c(\pi)$ for both $\sigma_{max}$ and  $\Delta \sigma$ are levelled. 

\begin{figure}[tb]

\includegraphics[trim = 5mm 0mm 18mm 0mm,clip,width=6.25cm, height=4.5cm]{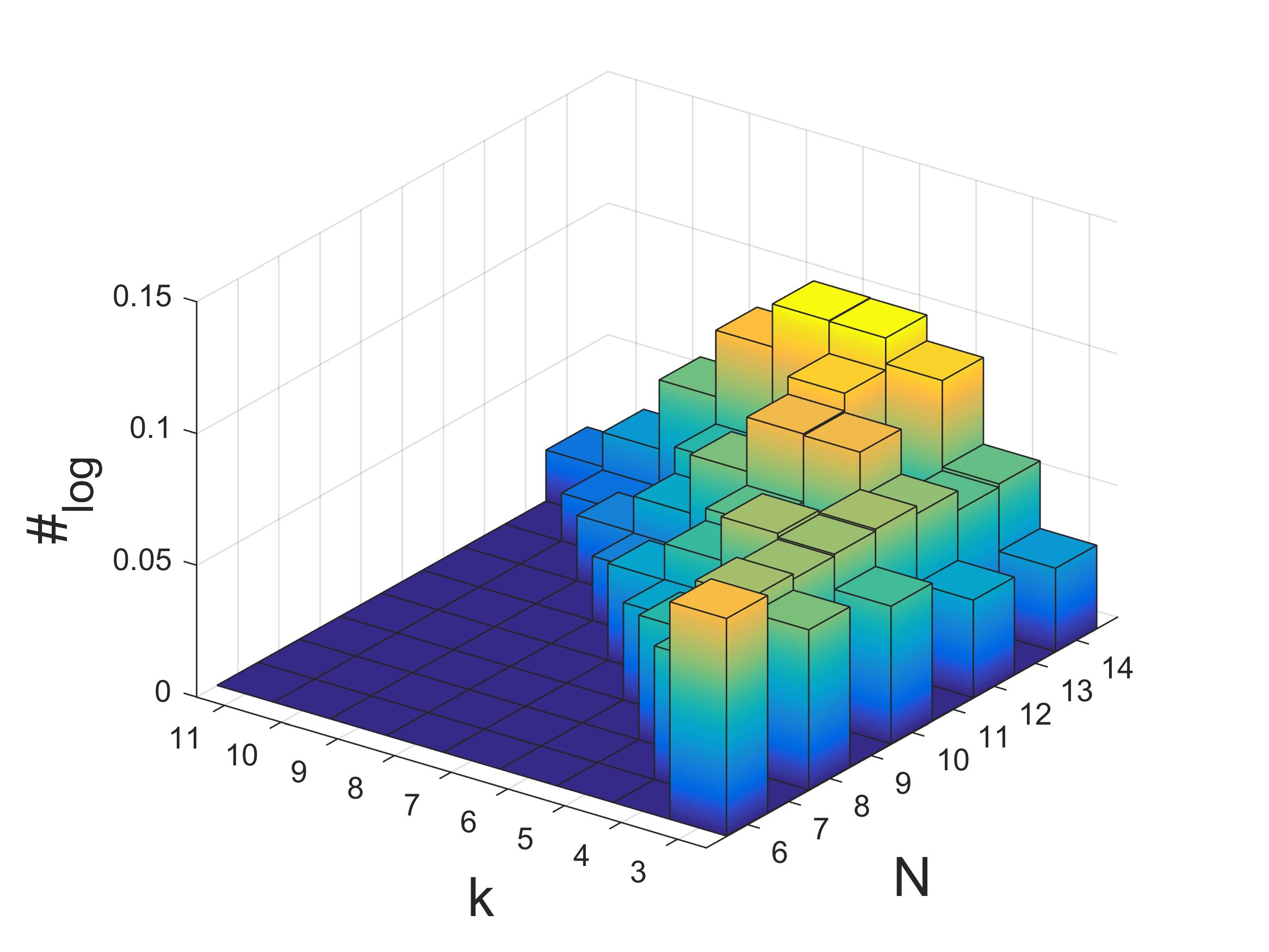}
\includegraphics[trim = 5mm 0mm 18mm 0mm,clip,width=6.25cm, height=4.5cm]{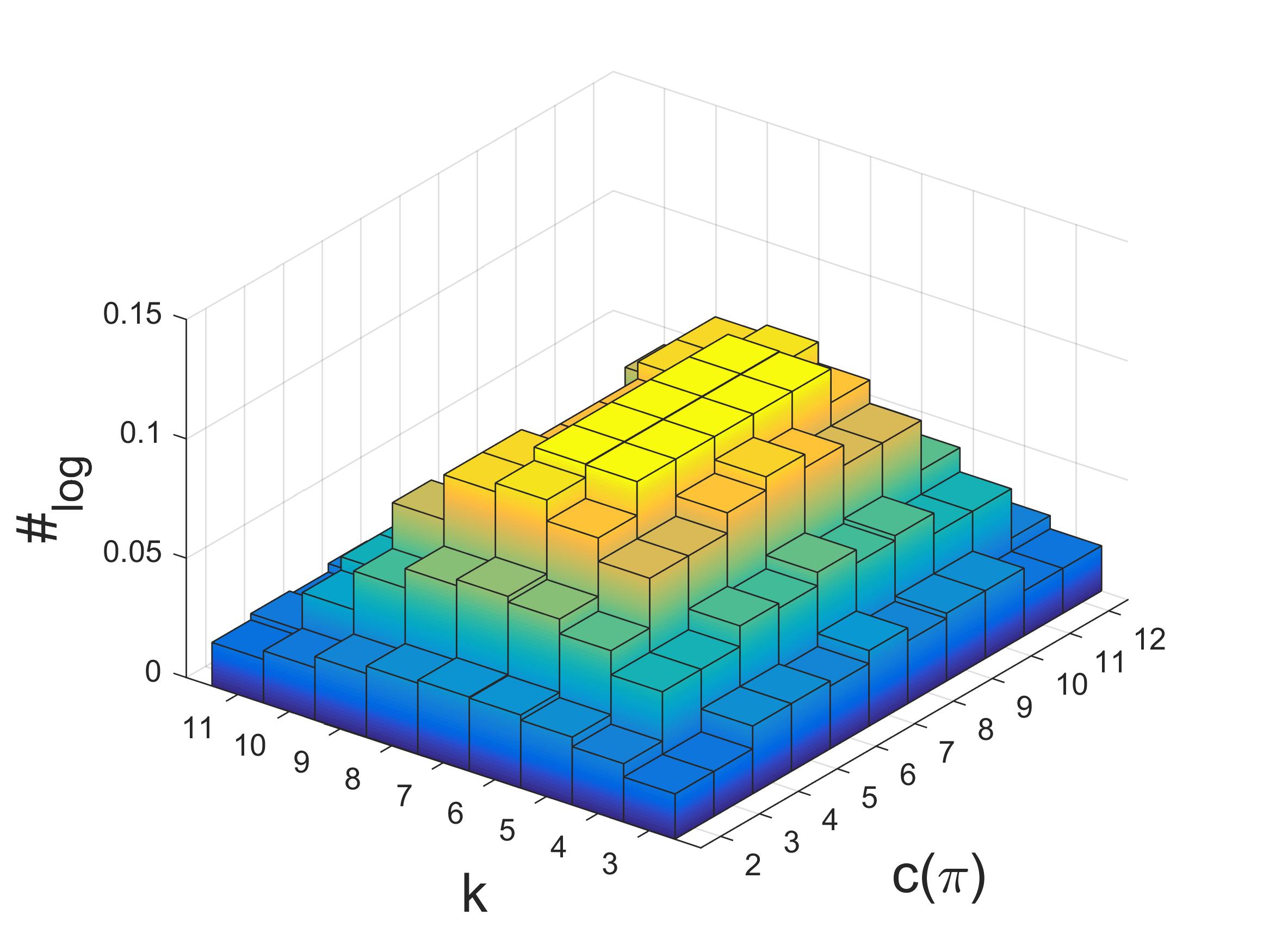}

\hspace{1cm}(a)  \hspace{5cm} (b) $N=14$

\caption{The quantity $\#_{log}=- \frac{1}{N^2} \log \left(\frac{\#_{\sigma_{max}}}{(4k-1/4k^2)\mathcal{L}_k(N)} \right)$ relating the number of $\sigma_{max}$--graphs, $\#_{\sigma_{max}}$, to the number of regular graphs $\mathcal{L}_k(N)$ for players $N$, coplayers $k$ and cooperators $c(\pi)$}
\label{fig:sig_log1}
\end{figure}

Apart from the numerical values of the maximal structure coefficients $\sigma_{max}$ and their relations to the number of players $N$, coplayers $k$ and cooperators $c(\pi)$, it is also interesting to know for which of the $\mathcal{L}_k(N)$ graphs the maximal values occurs. We call the graphs for which this happens the $\sigma_{max}$--graphs. Their number is $\#_{\sigma_{max}}$. Tab. \ref{tab:graphs1} give the number of $\sigma_{max}$--graphs, $\#_{\sigma_{max}}$, for all $N$ and $k$ considered here, see also Appendix 3 for some examples of $\sigma_{max}$--graphs. If we compare these numbers with the total number  $\mathcal{L}_k(N)$ of $k$--regular graphs on $N$ vertices, see Tab.  \ref{tab:graphs}, we observe that $\mathcal{L}_k(N)$ grows much faster than $\#_{\sigma_{max}}$. In other words, the   $\sigma_{max}$--graphs become rare as $N$ increases. Fig. \ref{fig:sig_log1} shows the quantity $\#_{log}=- \frac{1}{N^2} \log \left(\frac{\#_{\sigma_{max}}}{(4k-1/4k^2)\mathcal{L}_k(N)} \right)$ over $N$ and $k$ (Fig. \ref{fig:sig_log1}a), and over $c(\pi)$ and $k$ for $N=14$ (Fig. \ref{fig:sig_log1}b). We may conclude that as a rough approximation the ratio $\frac{\#_{\sigma_{max}}}{\mathcal{L}_k(N)}$  falls exponentially in $N$ and polynomially in $k$ for $k \approx N/2$ and $N$ getting larger. Furthermore, observe from Fig.  \ref{fig:sig_log1}b that for small and large values of the number of cooperators $c(\pi)$ there is a larger number of graphs that are $\sigma_{max}$--graphs. The $\sigma_{max}$--graphs become rarer for $c(\pi) \approx N/2$, for which but $\sigma_{max}$ is largest.

\begin{table}
\caption{The numbers $\#_{\sigma_{max}}$ of graphs with maximal $\sigma_{max}$ for all regular graphs with $\mathcal{L}_k(N)>1$ and $6 \leq N \leq 14$.  }
\label{tab:graphs1}
\center
\begin{tabular}{|c|c|c|c|c|c|c|c|c|c|c|}

\hline
 $\: {}_k \: \backslash \: {}^N$ & 6 & 7 & 8 & 9& 10 & 11 & 12 & 13 & 14 \\
 \hline
 3 & 1 & 0 & 1 &0 & 1 & 0 & 4 & 0 & 10 \\
 \hline
 4 & 0& 2 & 1 & 1 &1 & 1 & 2 & 10 & 14 \\
 \hline
 5 & 0 & 0 & 2 & 0 & 1 & 0 & 1 & 0 & 1 \\
 \hline
 6 & 0 & 0 & 0 & 3 & 2 & 5 & 1 & 2 & 1 \\
 \hline
 7 &0&0&0&0 & 2 & 0 & 4 & 0 & 1 \\
 \hline
 8 &0&0&0&0 & 0 & 5 & 6 & 49	& 4 \\
 \hline 
 9 &0&0&0&0& 0 & 0 & 4 & 0 & 14 \\
 \hline
 10 &0&0&0&0 & 0 & 0 & 0 & 7 & 14 \\ 
 \hline
 11 &0&0&0&0& 0 & 0& 0&0&4 \\

\hline

\end{tabular}
\end{table}

\begin{figure}[tb]
\includegraphics[trim = 5mm 0mm 18mm 0mm,clip,width=6.25cm, height=4.5cm]{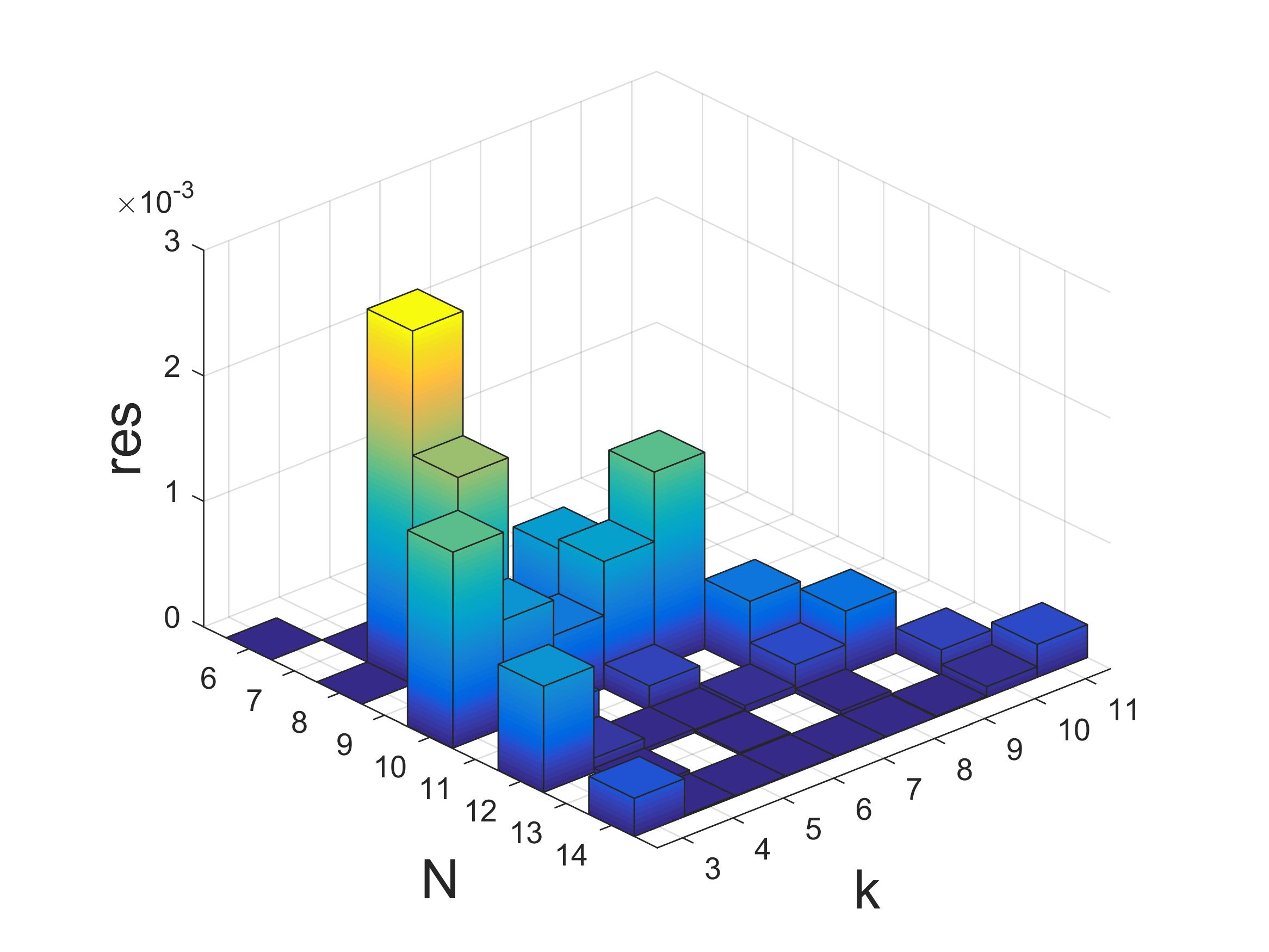}

\hspace{1cm}(a)  

\includegraphics[trim = 5mm 0mm 18mm 0mm,clip,width=6.25cm, height=4.5cm]{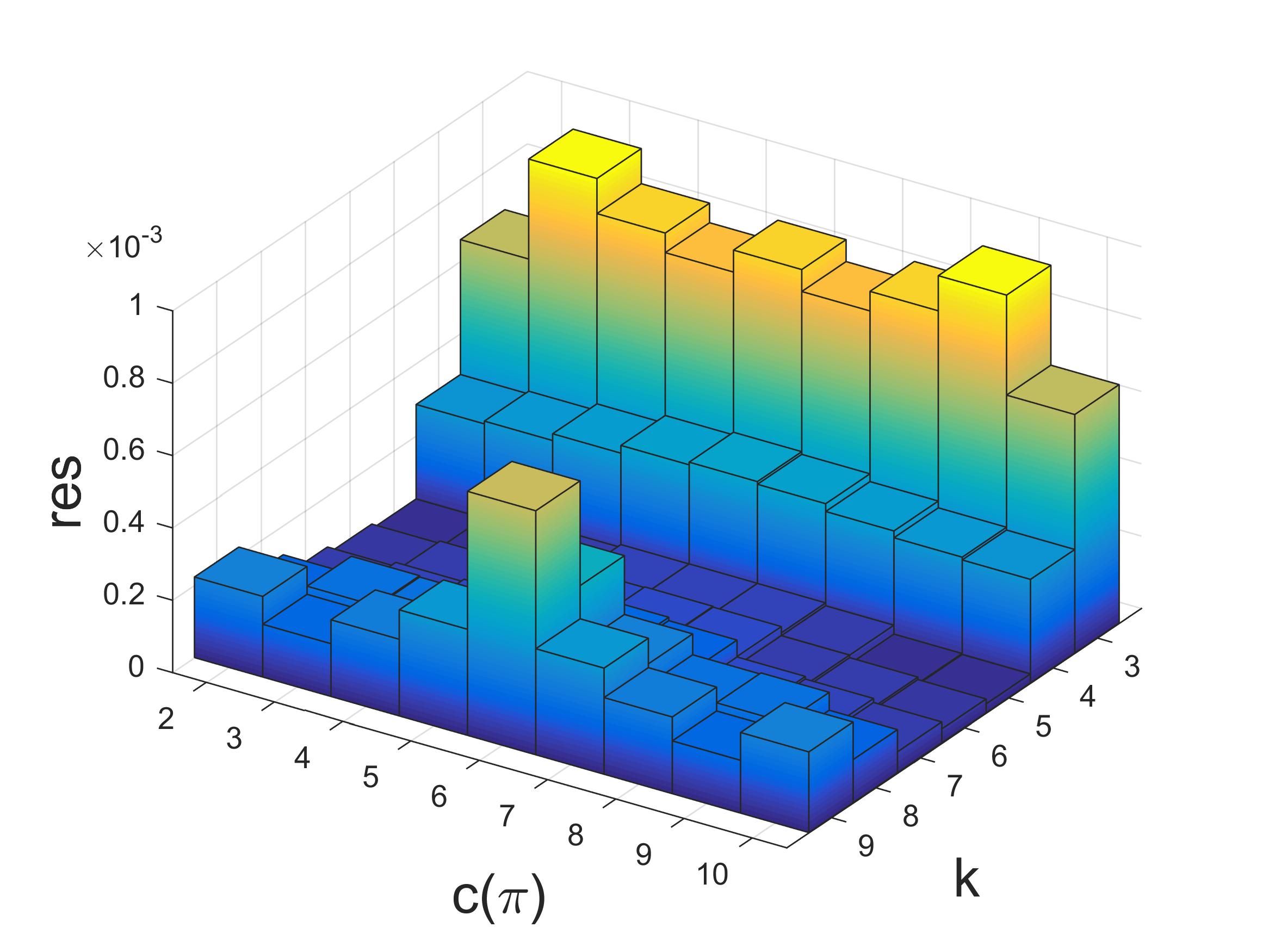}
\includegraphics[trim = 5mm 0mm 18mm 0mm,clip,width=6.25cm, height=4.5cm]{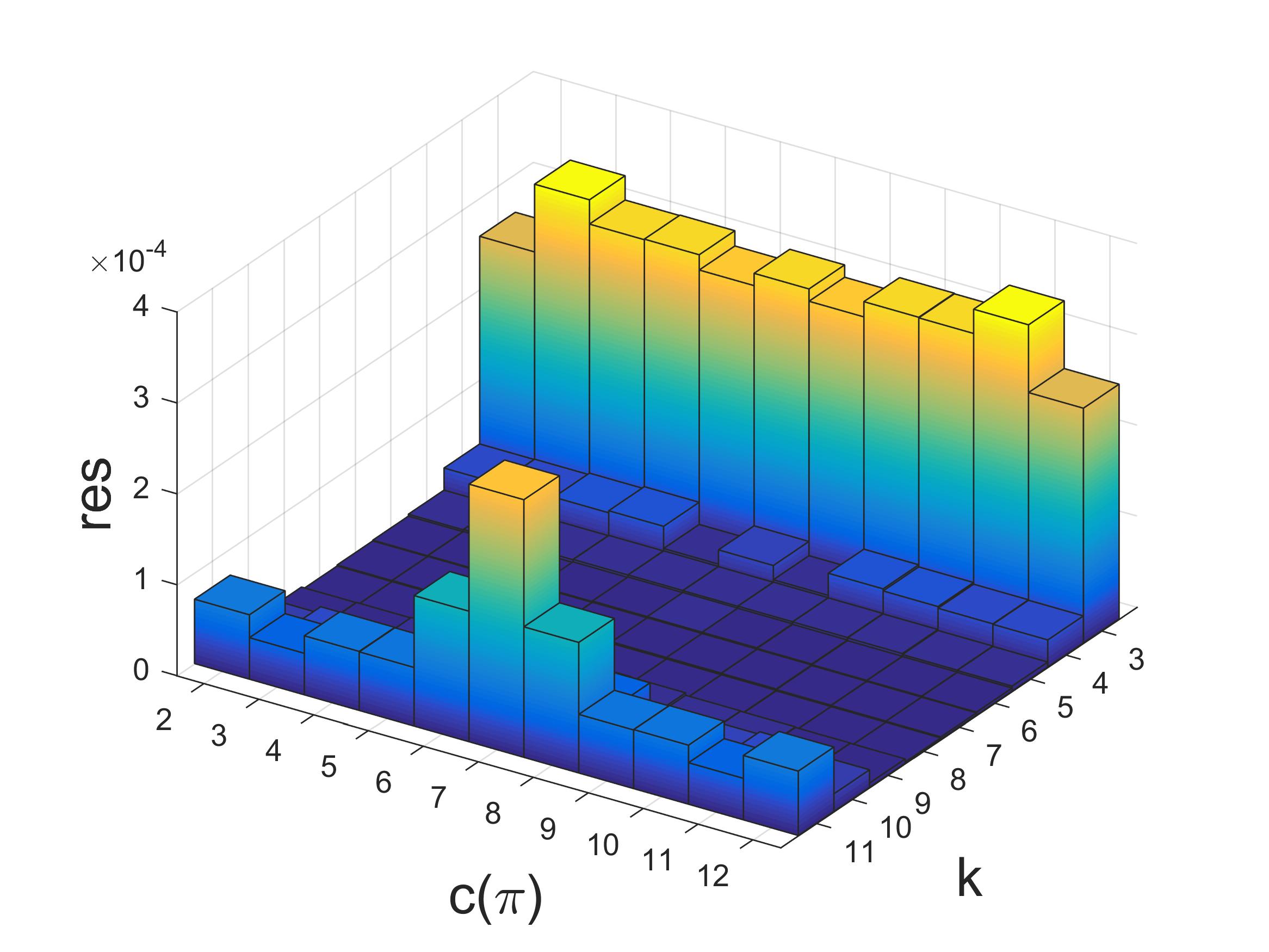}

\hspace{1cm}(b) $N=12$  \hspace{5cm} (c) $N=14$

\caption{Residual error $\text{res}$ according to Eq. (\ref{eq:res_error}) over $N$, $k$ and $c(\pi)$.}
\label{fig:sig_res}
\end{figure}

\subsection{Relationships between structure coefficients and graph cycles }

Recently, Giscard et al.~\cite{gis19} proposed an algorithm  to count efficiently the number of cycles with length $\ell$ in a graph: $\mathcal{C}_\ell(N,k)$ with $3 \leq \ell \leq N$. Thus, it is feasible to count $\mathcal{C}_\ell(N,k)$ for all $\mathcal{L}_k(N)$ regular graphs with $N\leq 14$, as given in Tab. \ref{tab:graphs}. As an example see  Fig.  \ref{fig:graph_6_3} with the count  $\mathcal{C}_\ell(6,3)$, $\ell=\{3,4,5,6\}$,  for the $\mathcal{L}_3(6)=2$ graphs with $N=6$ and $k=3$. The following discussion is based on
taking into account these numerical results. 

\begin{figure}[tb]

\includegraphics[trim = 55mm 140mm 120mm 20mm,clip,width=7.5cm, height=6cm]{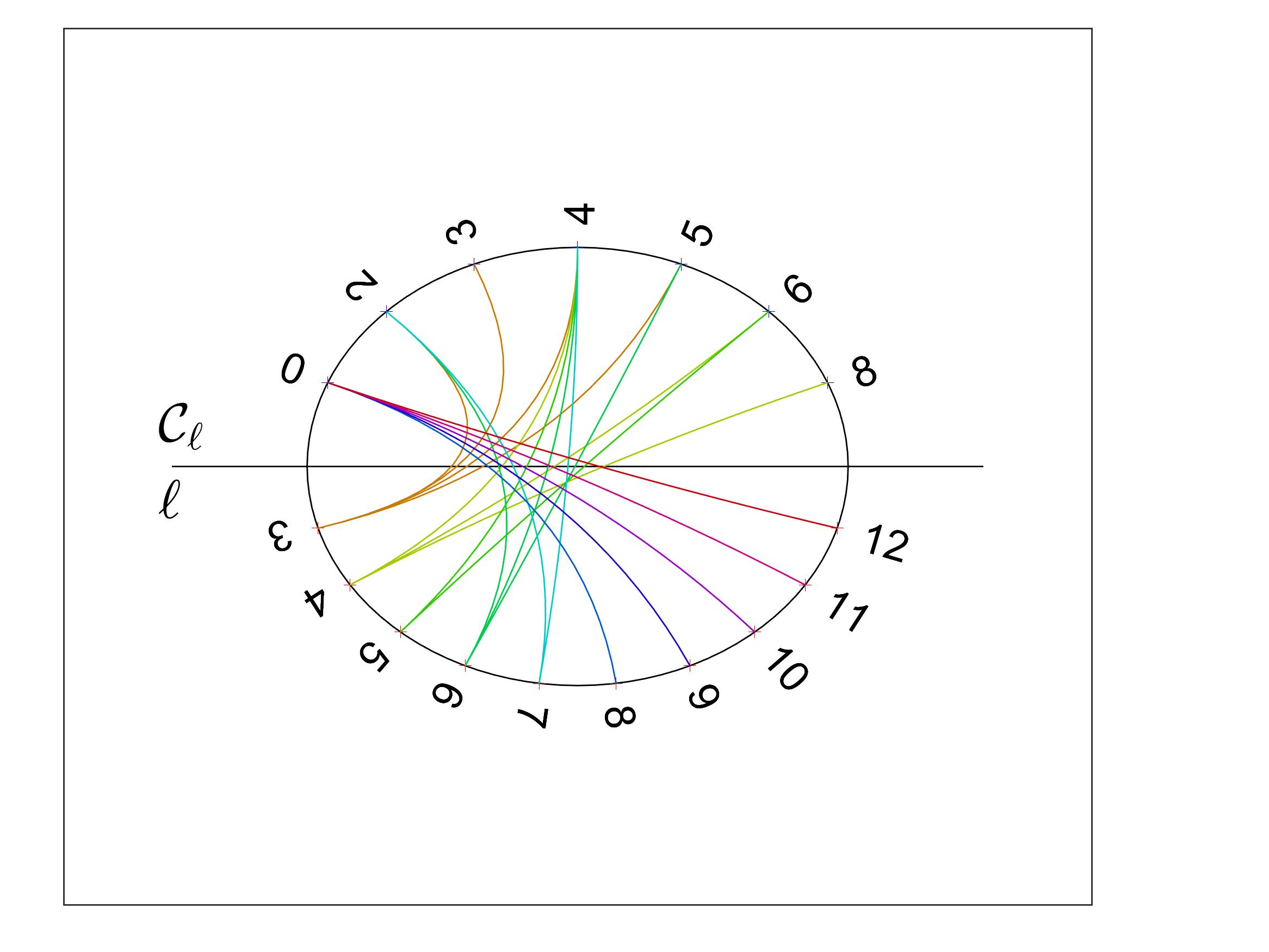}
\includegraphics[trim = 55mm 140mm 120mm 20mm,clip,width=7.5cm, height=6cm]{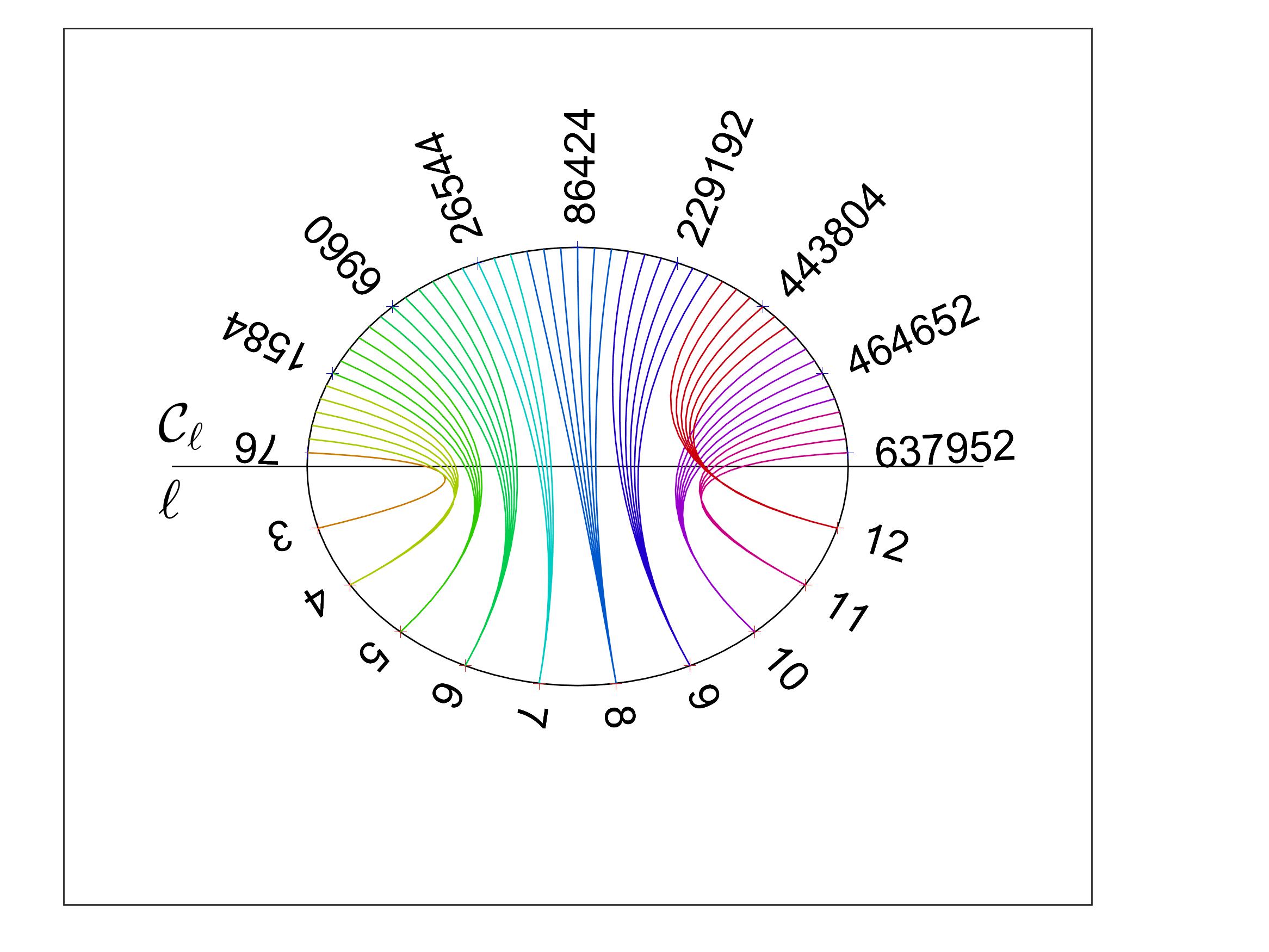}

\hspace{1cm}(a) $N=12$, $k=3$   \hspace{5cm} (b) $N=12$, $k=8$ 

\includegraphics[trim = 55mm 140mm 120mm 20mm,clip,width=7.5cm, height=6cm]{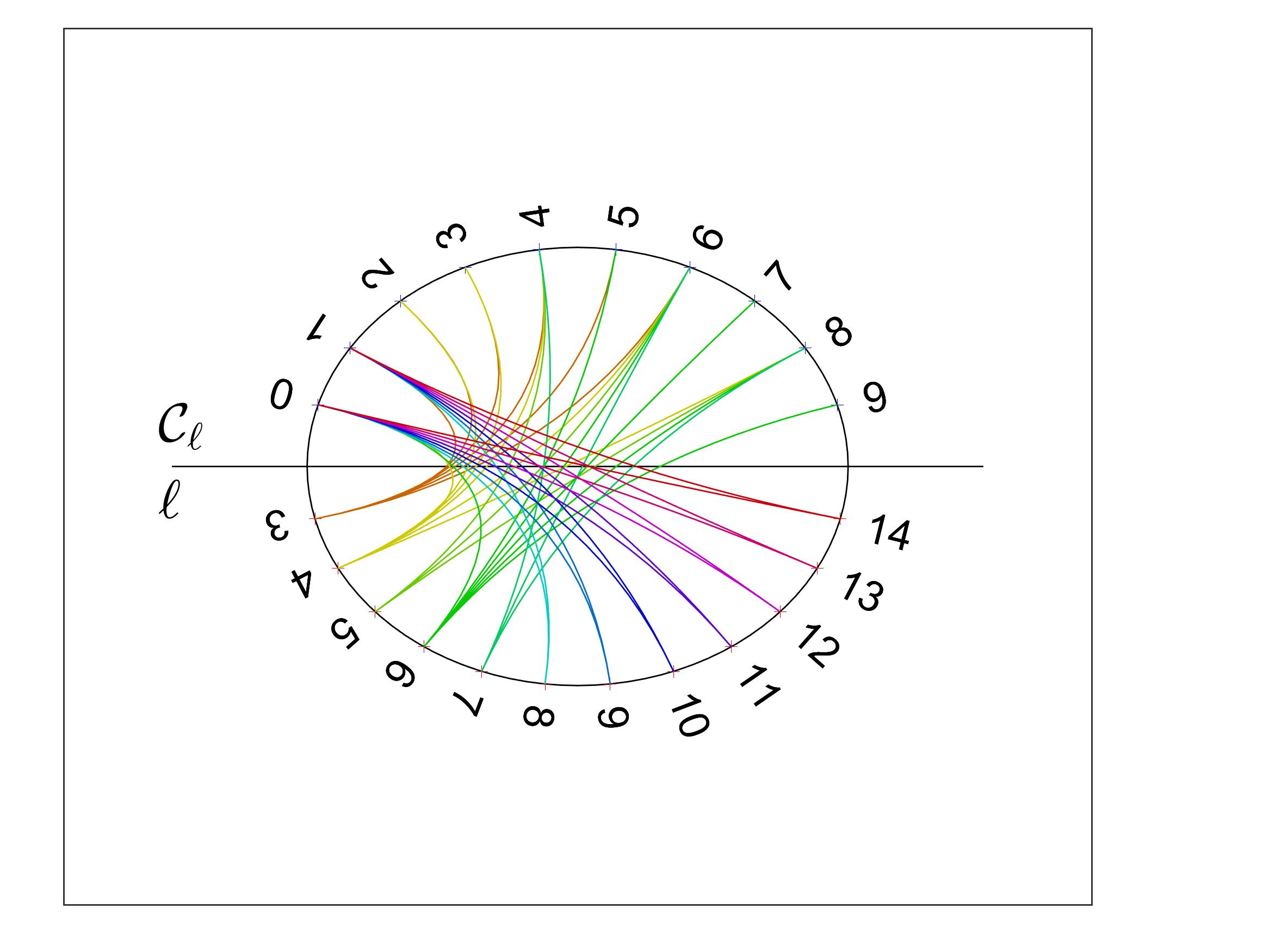}
\includegraphics[trim = 55mm 140mm 120mm 20mm,clip,width=7.5cm, height=6cm]{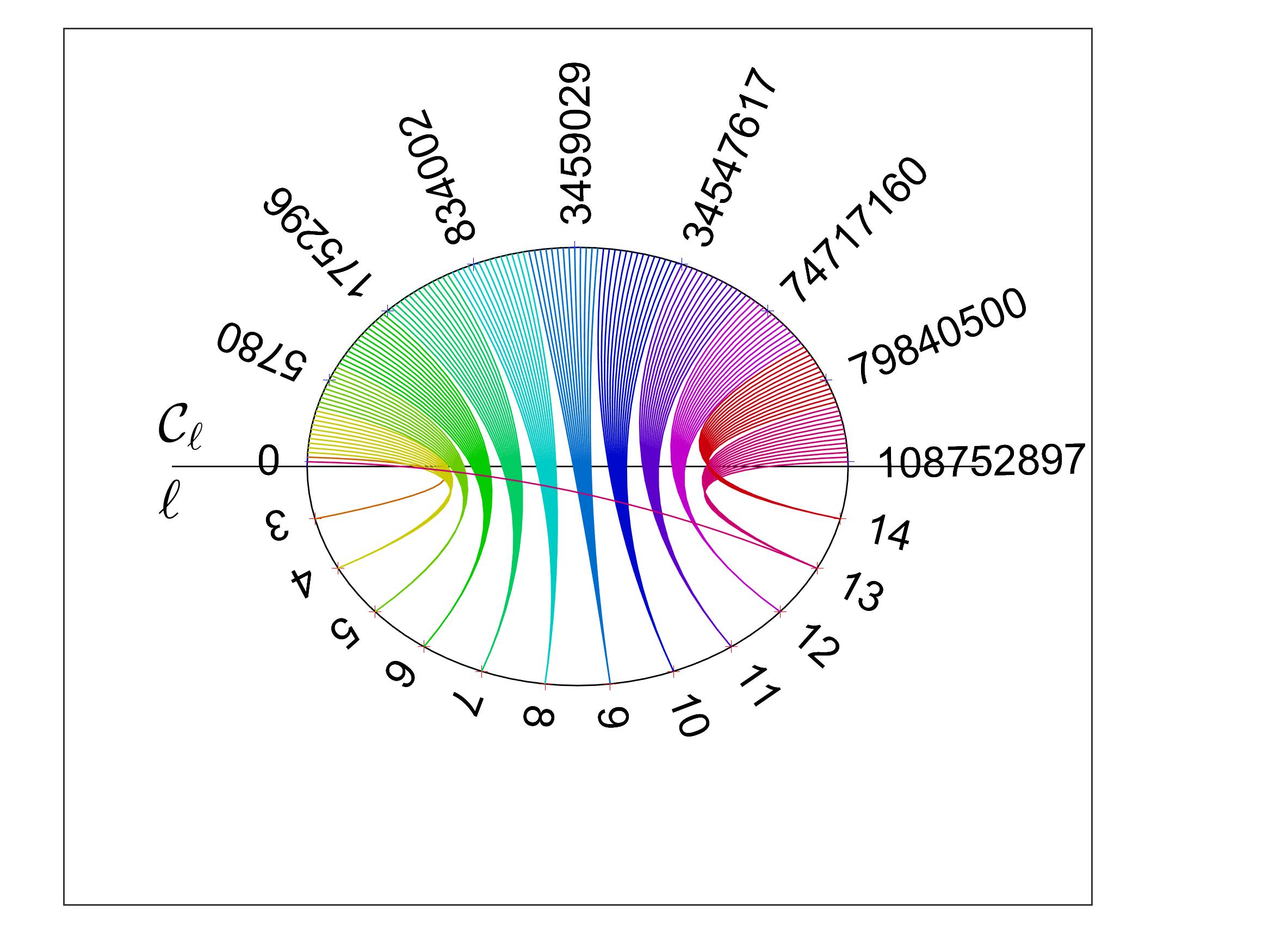}

\hspace{1cm}(c) $N=14$, $k=3$  \hspace{5cm} (d) $N=14$, $k=10$

\caption{Examples of schemaballs of $\sigma_{max}$--graphs.}
\label{fig:balls}
\end{figure}

In the previous section, it was shown that the maximal structure coefficients vary over interaction networks modelled as regular graphs, even if the number of players, coplayers and cooperators is constant. Thus, it appears reasonable to assume that some features of the graphs may be associated with these differences. In the following, results are presented in support for an approximately linear relationship between the number of graph cycles with certain length   and the maximal structure coefficients.   Two previous results can be interpreted as to point at the validity of such a relationship between the number of graph cycles  and fixation properties. A first is from evolutionary games on lattice grids~\cite{hau01,hau04,lang08,page00}. For these games, it has been shown that clusters of cooperators have a higher fixation probability than cooperators that are widely distributed on the grid. The location of the cluster on the grid does not matter. As lattice grids can be described by regular graphs (a Von Neumann neighborhood is a 4--regular graph, a Moore neighborhood a 8--regular graph) clusters imply short and closed paths between the nodes of the grid. Furthermore, the grid means an abundance of cycles with even cycle length. A second result is that between the structure coefficients and the path length between the cooperators there is a strong negative correlation~\cite{rich19b}.  Cooperator path length is defined as the path length averaged over all pairs of cooperators on the evolutionary graph. If there are more than two cooperators, the cooperator path length has particularly small values if the cooperators cluster next to each other and are linked by loops. Thus, small values of the cooperator path length correspond with the abundance of cycles of certain length.

\begin{figure}[tb]

\includegraphics[trim = 55mm 140mm 120mm 20mm,clip,width=7.5cm, height=6cm]{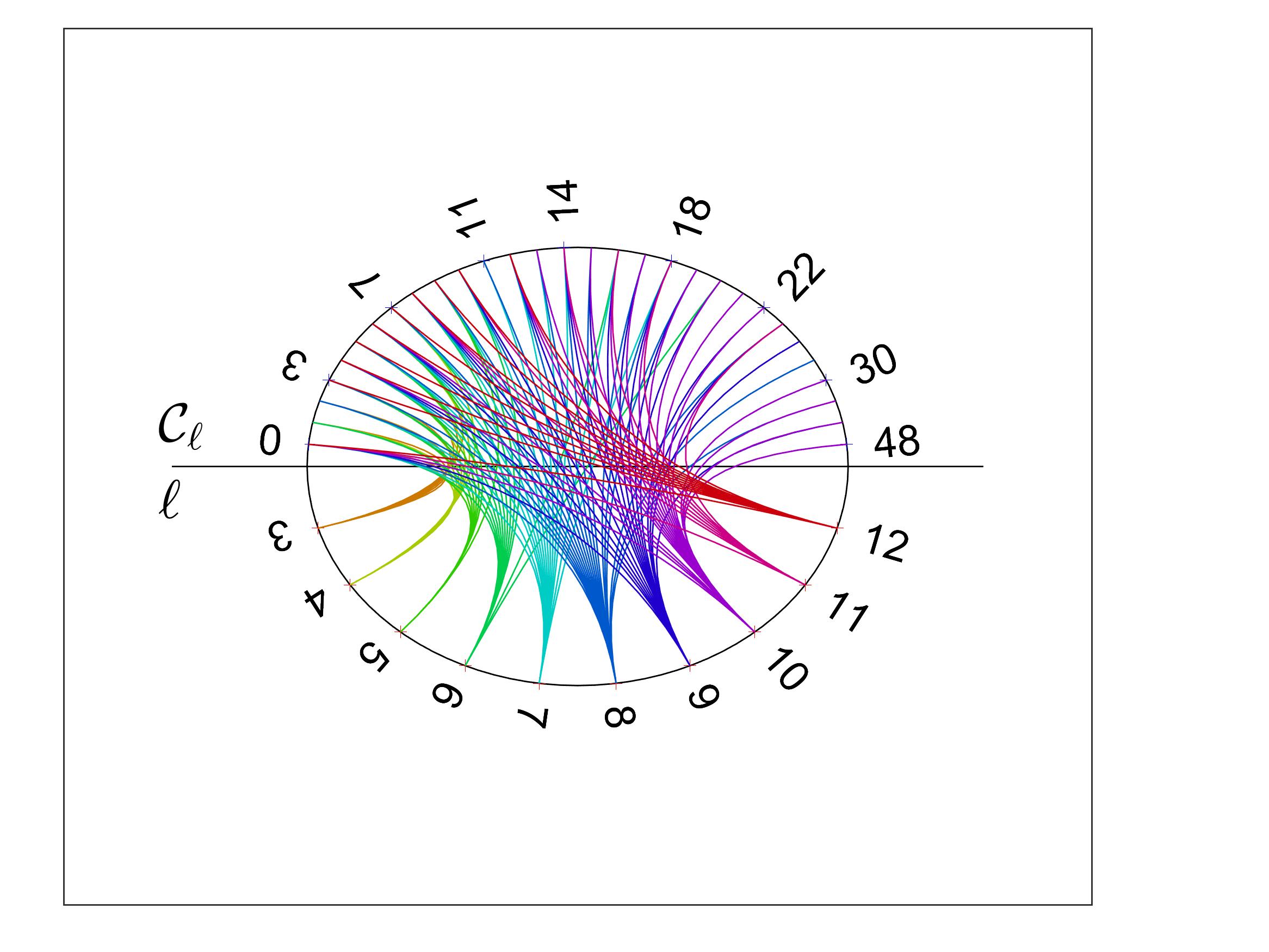}
\includegraphics[trim = 55mm 140mm 120mm 20mm,clip,width=7.5cm, height=6cm]{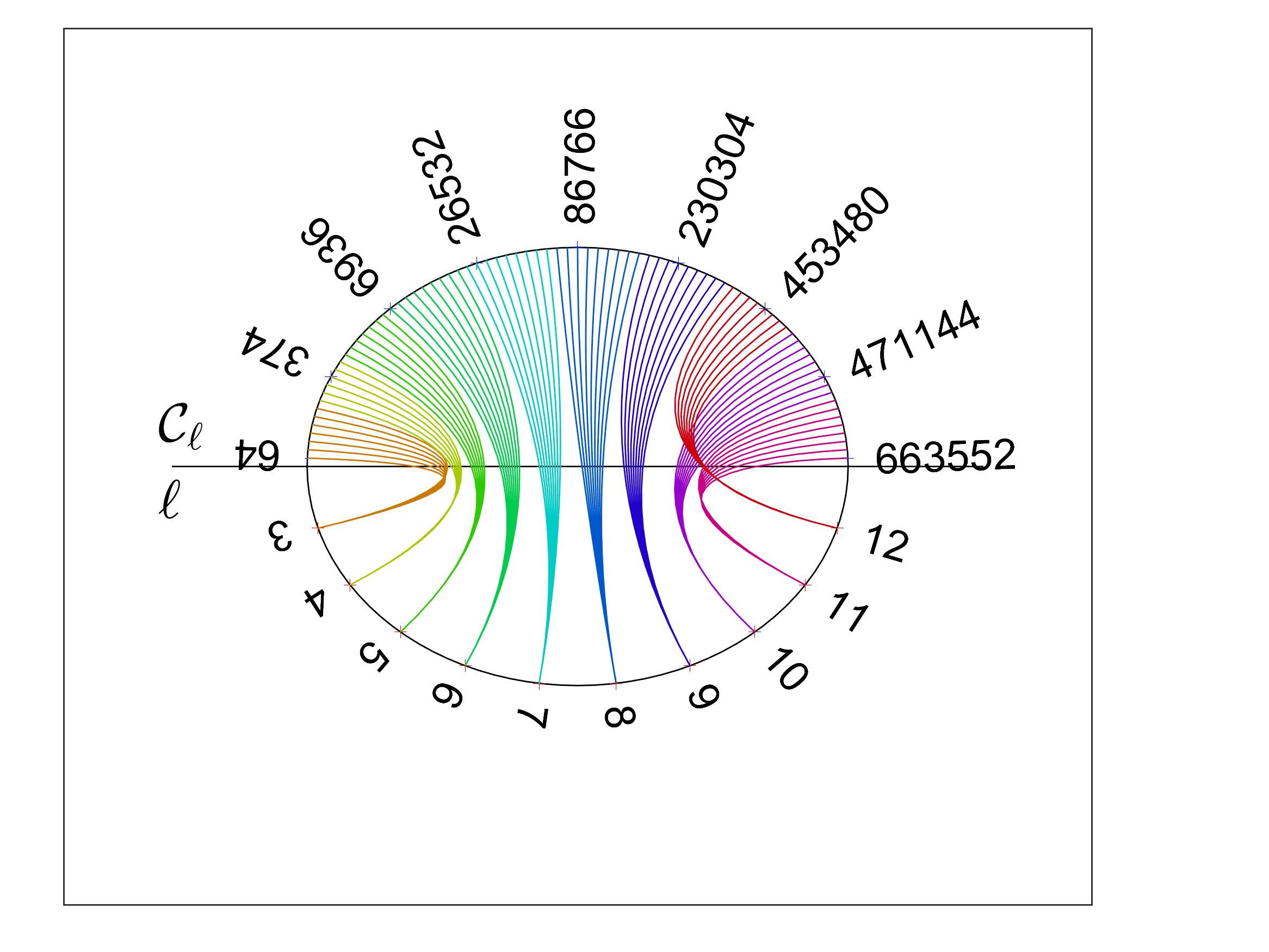}

\hspace{1cm}(a) $N=12$, $k=3$   \hspace{5cm} (b) $N=12$, $k=8$ 

\includegraphics[trim = 55mm 140mm 120mm 20mm,clip,width=7.5cm, height=6cm]{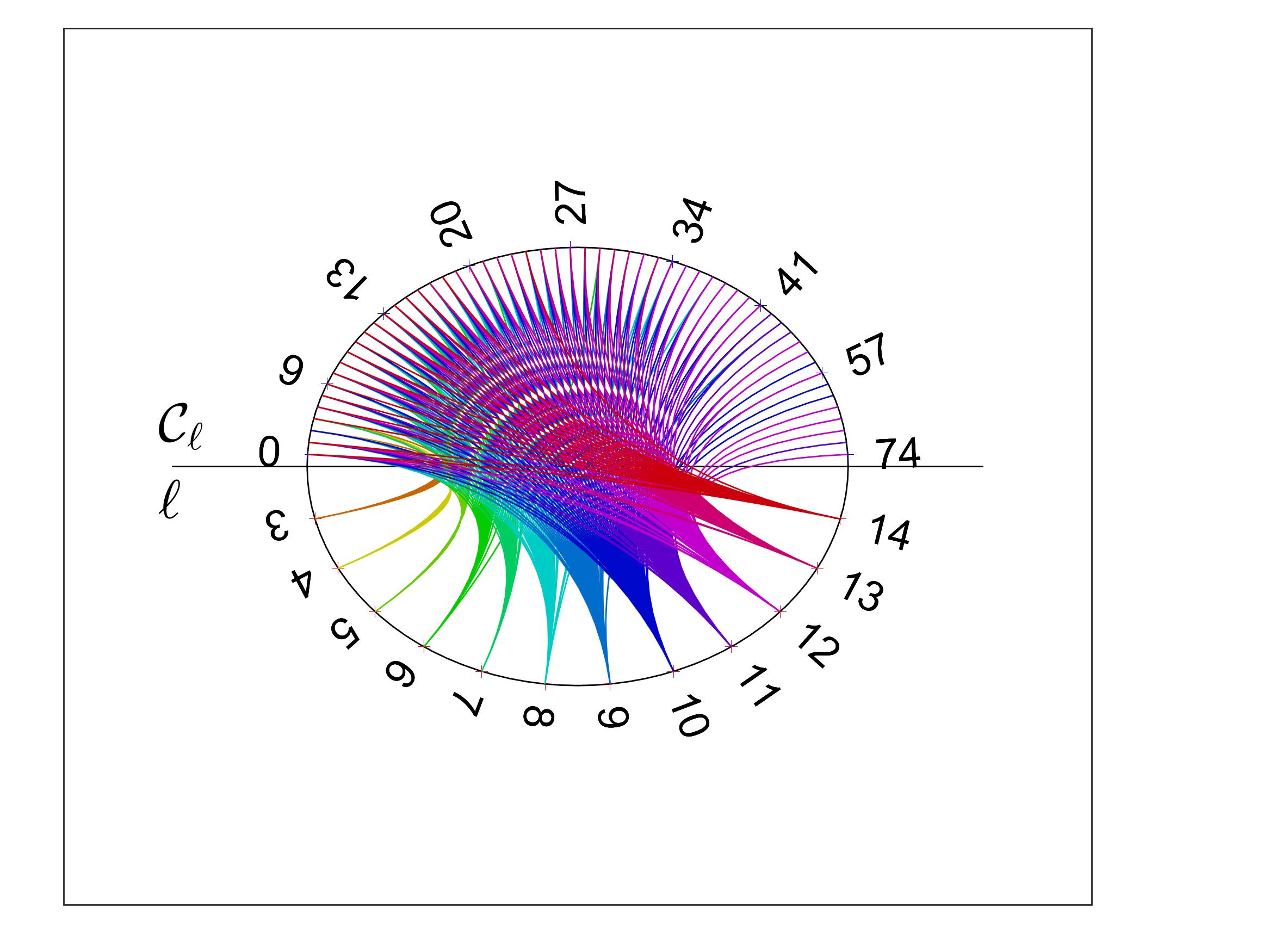}
\includegraphics[trim = 55mm 140mm 120mm 20mm,clip,width=7.5cm, height=6cm]{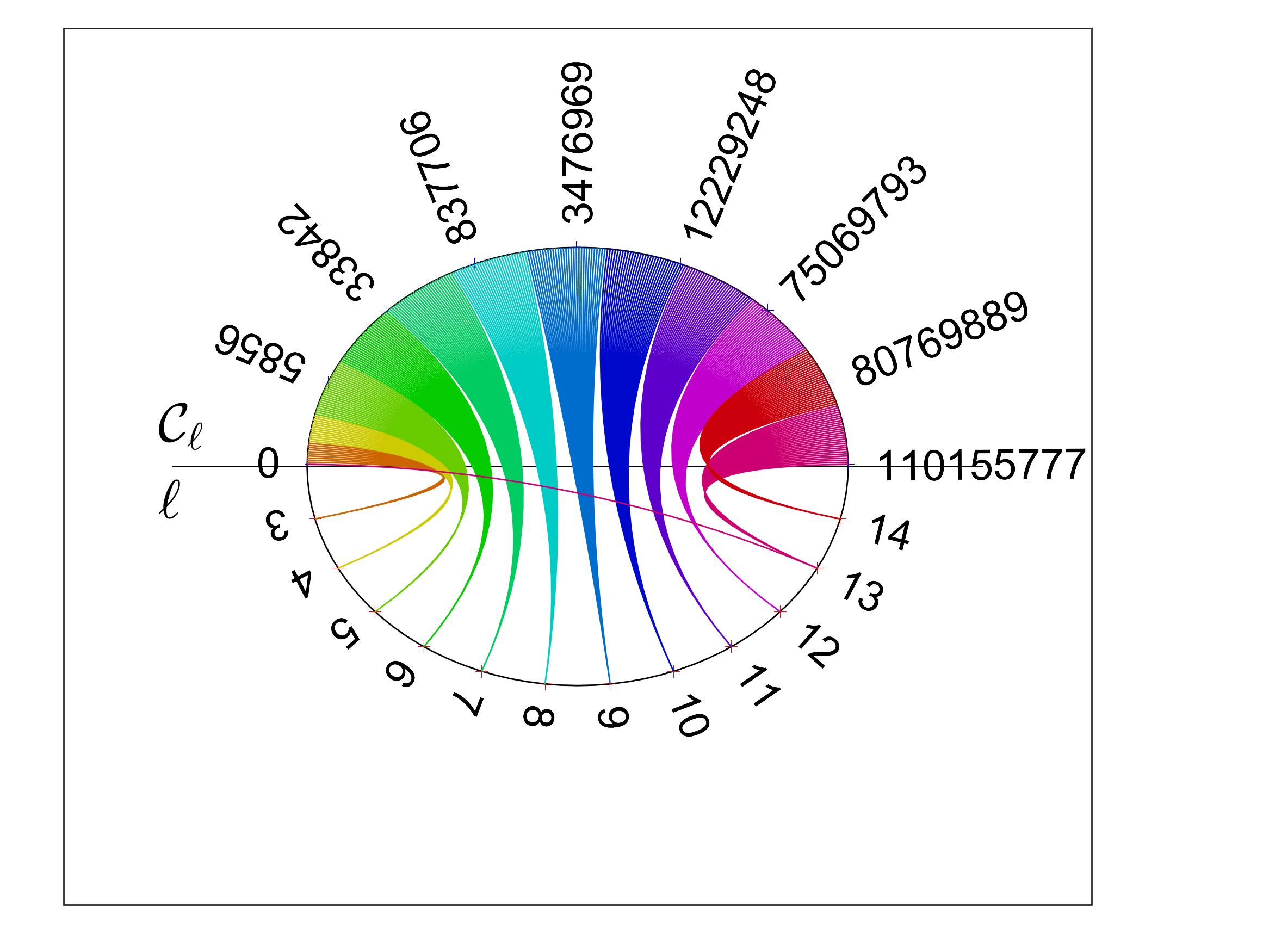}

\hspace{1cm}(c) $N=14$, $k=3$  \hspace{5cm} (d) $N=14$, $k=10$

\caption{Examples of schemaball of $\sigma_{min}$--graphs.}
\label{fig:balls1}
\end{figure} 

As there are $\mathcal{L}_k(N)$ regular graphs for a given $N$ and $k$, we obtain $\mathcal{L}_k(N)$  maximal structure coefficients $\sigma_{max_i}$, $i=1,2,\ldots\mathcal{L}_k(N)$ together with the same count of cycle length $\mathcal{C}_{\ell_i}(N,k)$. Thus, we may assume for each $i$ a linear relationship $\sigma_{max_i}=\mathcal{C}_{\ell_i}(N,k) x +\epsilon_i$ for some variables $x$ with an error term $\epsilon_i$. 
To test the validity of this linear relationship, we calculate the residual  error 
\begin{equation}\text{res}=\frac{1}{\mathcal{L}_k(N)} \|  \mathbfcal{C}_\ell x^* - \boldsymbol\sigma_{max} \|,  \label{eq:res_error} \end{equation}
where $\mathbfcal{C}_\ell$ comprises of all $\mathcal{L}_k(N)$ cycle length  $\mathcal{C}_{\ell_i}(N,k)$ and $\boldsymbol\sigma_{max}$ contains all $\mathcal{L}_k(N)$  structure coefficients $\sigma_{max_i}$ for a given $N$ and $k$. The variable $x^*$ is the solution 
of the non--negative least square problem 
\begin{equation}x^*=\arg \min_x \|  \mathbfcal{C}_\ell x - \boldsymbol\sigma_{max} \|. \end{equation}
As the length of $x^*$ varies with varying $\mathcal{L}_k(N)$, the residual error in (\ref{eq:res_error}) is weighted by $\mathcal{L}_k(N)$ to make it comparable over all $N$ and $k$. Note that the residual error (\ref{eq:res_error}) gives equivalent results to the root--mean--square deviation, which is also sometimes used to measure the accuracy of a (linear) model. 
The results are given in Fig. \ref{fig:sig_res}. We see that the residual error $\text{res}$ is small for all $6 \leq N \leq 14$, $3 \leq k \leq N-3$ and gets even smaller for $N$ getting larger. Generally, the error $\text{res}$ is slightly larger for $k=3$ and $k=N-3$ than for intermediate values of $k$. This is also true for calculating $\text{res}$ for each number of cooperators $c(\pi)$, see Figs. \ref{fig:sig_res}b and \ref{fig:sig_res}c, which show the results for $N=12$ and $N=14$. For $N=14$ the values of $\text{res}$ are generally smaller than for $N=12$ and the largest values of $\text{res}$ are obtained for small and large $k$ for all $c(\pi)$. To conclude we can observe that the results for the residual error $\text{res}$ are generally very small, which is equivalent to saying that the error term $\epsilon_i$ in the assumed linear relationship $\sigma_{max_i}=\mathcal{C}_{\ell_i}(N,k) x +\epsilon_i$ has an expected value $\mathbb{E}(\epsilon_i) \approx 0$. Thus, there is some justification to observe that  between the maximal structure coefficients $\sigma_{max_i}$ and the cycle count  $\mathcal{C}_{\ell_i}(N,k)$ there is an approximately linear relationship.

Finally, another aspect of
the interplay between graph structure and fixation properties should be highlighted. To begin with, we analyze the  cycle count  $\mathcal{C}_{\ell}(N,k)$ of $\sigma_{max}$--graphs, which are those graphs among the $\mathcal{L}_k(N)$ regular graphs that have maximal structure coefficients. Consider the example $N=12$ and $k=3$. There are $\mathcal{L}_3(12)=85$ graphs of which $\#_{\sigma_{max}}=4$ are $\sigma_{max}$--graphs, compare Tab. \ref{tab:graphs1} with Tab. \ref{tab:graphs}. For these 4 graphs we analyze how the count $\mathcal{C}_{\ell}(12,3)$ is distributed over $\ell=3,4,\ldots,12$. A possible way to visualize such an analysis  is based on schemaballs~\cite {krz04,rich19a}, see Fig. \ref{fig:balls}a.     In such a schemaball we draw Bezier curves connecting the count $\mathcal{C}_\ell(N,k)$ in the upper half of the ball with the associated cycle length $\ell$ in the lower half. The actual values of both $\ell$ and  $\mathcal{C}_\ell(N,k)$ are written on the ball.
 The curves are colored in such a way that equal values of the cycle length $\ell$ have the same (and specific) color, no matter to which cycle count $\mathcal{C}_\ell(N,k)$ they are belonging. The colors are selected  equidistant from a RGB color wheel. If there are several $\sigma_{max}$--graphs, as there are $\#_{\sigma_{max}}=4$ for $N=12$, $k=3$ in Fig. \ref{fig:balls}a, each graph has its own set of curves between $\ell$ and $\mathcal{C}_\ell$. The schemaball thus contains all of them, which means there may be curves between the same  value of $\ell$ and several $\mathcal{C}_\ell$ (and vice versa). For instance, in Fig. \ref{fig:balls}a showing the schemaball for $N=12$ and $k=3$, we see that for $\ell=3$, which is cycles of length 3, also known as triangle, we find connection to $\mathcal{C}_3(13,3)=(2,3,4,5)$. This means each of the $\#_{\sigma_{max}}=4$ graphs has triangle, one has 2 of them, another one has 3, still another one has 4 and the final one has 5 triangle. 
 
 From the visualization using a schemaball it can be immediately seen that for $N=12$ and $k=3$ small cycles lengths, that is $\ell=\{3,4,\ldots,7\}$, have generally a count 
$\mathcal{C}_\ell(12,3)>0$. For large cycle lengths, that is 
$\ell=\{8,9,\ldots,12\}$, we have $\mathcal{C}_\ell(12,3)=0$. For $N=14$ and $k=3$, see Fig. \ref{fig:balls}c, we get very similar results. By contrast, for larger $k$, not only the cycle count $\mathcal{C}_\ell(N,k)$ is much higher than for lower $k$, but also the distribution over cycles lengths $\ell$ is quite different, see the examples $N=12$, $k=8$, Fig. \ref{fig:balls}b and $N=14$, $k=10$, Fig. \ref{fig:balls}d. Here, small as well as large cycle lengths $\ell$ have a substantial count   $\mathcal{C}_\ell(N,k)$. Moreover, every cycles length $\ell$ is connected to a distinct interval of $\mathcal{C}_\ell(N,k)$. This means that the $\sigma_{max}$--graphs have very similar counts $\mathcal{C}_\ell(N,k)$ for each $\ell$. These properties becomes even more clear if we additionally consider the schemaballs for $\sigma_{min}$--graphs, which are the graphs with minimal structure coefficients see Fig. \ref{fig:balls1} for the same examples as Fig. \ref{fig:balls}. 
Not only there are more $\sigma_{min}$--graphs than $\sigma_{max}$--graphs, (for instance 77 vs. 4 for $N=12$, $k=3$, or 359 vs. 6 for $N=12$, $k=8$), the balls for small $k$ look very different, compare Figs. \ref{fig:balls1}a and \ref{fig:balls1}c with Figs. \ref{fig:balls}a and \ref{fig:balls}c.  For the $\sigma_{min}$--graphs and small $k$ even large cycle length $\ell$  have a substantial count $\mathcal{C}_\ell(N,k)$. The count is actually much higher, which means that $\sigma_{min}$--graphs have generally more cycles of a given length than $\sigma_{max}$--graphs.  On the other hand, for large $k$ the differences are rather marginal. The only difference is that the schemaballs are more dense, which means that $\sigma_{min}$--graphs have more different counts for a given cycle length than $\sigma_{max}$--graphs. For the other tested number of players $N \leq 14$ similar results   are obtained as shown in Figs. \ref{fig:balls} and \ref{fig:balls1}. We next discuss some implications of these results on the evolution of cooperation on regular evolutionary graphs.

\section{Discussion and Conclusions}
In this paper structure coefficients $\sigma(\pi,\mathcal{G})$ introduced by Chen et al.~\cite{chen16} (see~\cite{rich19a,rich19b} for further analysis) are studied for all regular interaction graphs with $N\leq 14$ players and $3 \leq k \leq N-3$ coplayers. These structure coefficients provide   a simple condition   connecting long--term prevalence of cooperation with the values of the payoff matrix (\ref{eq:payoff}), the structure of the evolutionary graph $\mathcal{G}$ and the arrangement of any number of cooperators and defectors on this graph, which is expressed by the configuration $\pi$. Cooperation is favored for weak selection and a configuration $\pi$ on a graph $\mathcal{G}$ if \begin{equation}\sigma(\pi,\mathcal{G})>\frac{c-b}{a-d}. \label{eq:cond1}\end{equation}
For $\sigma(\pi,\mathcal{G})<1$, the game favors the evolution of spite, which can be seen as a sharp opposite to cooperation.  For $\sigma(\pi,\mathcal{G})=1$, the condition (\ref{eq:cond1}) matches the standard condition of risk--dominance. 
For $\sigma(\pi,\mathcal{G})>1$, the diagonal elements of the payoff matrix (\ref{eq:payoff}), $a$ and $d$, are more critical than the off--diagonal elements, $b$ and $c$, for determining which strategy is favored. For instance,  cooperation can be favored in the Prisoner's Dilemma game, which is specified by $c>a>d>b$. 
The condition (\ref{eq:cond1}) implies that a larger value of $\sigma(\pi,\mathcal{G})$ still allows cooperation to emerge if $a-d$ is small (or $c-b$ is large). For the Stag Hunt game (Coordination game), characterized by $a>c \geq d>b$, the condition   $\sigma(\pi,\mathcal{G})>1$ means to favor a Pareto--efficient strategy ($a>d$) over a risk--dominant strategy ($a+b<c+d$). Again, a larger value of $\sigma(\pi,\mathcal{G})$ tolerates a smaller Pareto--efficiency $a-d$.
Put differently, cooperation is favored even if the difference between reward and punishment is rather low. 
A generalization of these discussions can be achieved by the universal scaling approach for payoff matrices that facilitates studying a continuum of social dilemmas~\cite{wang15}. According to this approach a larger value of $\sigma(\pi,\mathcal{G})$  implies a larger section of the parameter space spanned by gamble--intending and risk--averting dilemma strength~\cite{rich19c}. Based on this interpretation of the structure coefficient  $\sigma(\pi,\mathcal{G})$, we review the following major results of
the numerical experiments presented in Sec. 2.
\begin{itemize}

\item[a.] There is an approximately linear relationship between maximal structure coefficients and the count of cycles of the interaction graph with certain length. Moreover, the number of $\sigma_{max}$--graphs grows much slower for a rising number of players than the number of $k$--regular graphs on $N$ vertices.  Thus, graphs with maximal structure coefficients get rare for the number of players $N$ getting large.

\item[b.] The values of the structure coefficients are larger for a small number of coplayers, that is for graphs with a small degree, and maximal for $k=3$, which is cubic graphs, than for larger numbers of coplayers. This is also the case for the largest differance between maximal and minimal structure coefficients. Thus, for regular evolutionary graphs describing the interactions between players, the results for $N \leq 14$ players suggest  
that a smaller number of coplayers is particularly prone to promote cooperation if a favorable graph is selected. The selection of graphs does matter less for a larger number of coplayers. 
The $\sigma_{max}$--graphs with small numbers of coplayers $k$ not only have largest maximal structure coefficients, they are also characterized by the absence of cycles with a length above a certain limit, see examples in the collection of  $\sigma_{max}$--graphs  in Appendix 3.

\item[c.] 
 There are not only no long cycles in $\sigma_{max}$--graphs with small $k$. The graphs are also structured into blocks that are connected by cut vertices and/or hinge vertices. A cut vertex is a vertex whose removal disconnects the graph, while a hinge vertex is a vertex whose removal makes the distance longer between at least two other vertices of the graphs~\cite{chang97,ho96}. For instance, for $N=12$ and $k=3$, the vertices occupied by the players $\mathcal{I}_3$ and $\mathcal{I}_9$, see Fig. \ref{fig:graph_12_3}, are cut vertices, while for $N=10$ and $k=4$,  see Fig. \ref{fig:graph_10_x}b, the vertices occupied by the players $\mathcal{I}_5$ and $\mathcal{I}_6$ are  hinge vertices as their removal would make the distance between  $\mathcal{I}_4$ and  $\mathcal{I}_7$ longer.   The blocks are occupied by clusters of cooperators.  The clusters can be seen as to serve as a mutant family that invades the remaining graph. As vertices with players of opposing strategies are connected by cut and/or hinge vertices there is only a small number of (or even just a single) migration path for the cooperators and/or defectors.  A similar observation has been reported for evolutionary games on lattices grids~\cite{hau01,lang08}, see also the discussion in Sec. 2.2. To summarize: the results suggest that $\sigma_{max}$--graphs for small numbers of coplayers have some distinct graph--theoretical properties. Searching for these properties in a given graph may inform the design of interactions graphs that are either particularly prone to cooperation or particularly opposed to it.

\item[d.] The property of missing long cycles is also a possible explanation as to why regular graphs with small degree differ substantially from graphs with larger degree in terms of promoting cooperation in evolutionary games. A larger degree makes it impossible to have blocks that are connected by only a few edges. As the number of edges increases linearly with the degree by $kN/2$ and each vertex has the same number of edges, there is an ample supply on connections. These results imply that connectivity properties of the interaction graph play an important role in the emergence of cooperation. It may be interesting to see if  these
connectivity issues may possibly also show in algebraic graph measures, for instance algebraic connectivity expressed by the Fiedler vector.

\end{itemize}
The results given above show a clear dependency between the long--term prevalence of cooperation in evolutionary games on regular graphs and some of their graph--theoretical properties, which generally confirm previous findings on clusters of cooperators in games on lattice  
grids~\cite{hau01,hau04,lang08,page00}, on pairs of mutants on a circle graph ($k=2$)~\cite{xiao19}, and on short cooperator path lengths on some selected regular graphs with $N=12$ and $k=3$, among them the Frucht, the Tietze and the Franklin graph~\cite{rich19b}. However, apart from statements about the prevalence of cooperation there are also other quantifiers of evolutionary dynamics that are highly relevant. In other words, 
some of the difficulty in the given approach for evaluating the emergence of cooperation in evolutionary games on graphs arises from structure coefficients merely treating a comparison of fixation probabilities. The condition indicates that the fixation probability of cooperation is higher than the fixation probability of defection. This, however, does not entail the values of these probabilities.   However, structure coefficients can be calculated with polynomial time complexity~\cite{chen16}, while computing fixation probabilities is generally intractable due to an exponential time complexity~\cite{hinder16,ibs15,vor13}. In other words, by using the approach involving structure coefficients, we exchange computational tractability by obtaining just a comparison of fixation probabilities instead of their exact values.   Moreover,  apart from the difference in the information obtained, the variety in the descriptive power of the structure coefficients as compared to the fixation probabilities is salient in another way. Most likely, there is a rather complex relationship between structure coefficients and fixation probability, which is illustrated by the example of a single cooperator for which the structure coefficient does precisely not imply  unique values of the fixation probability of cooperation. For a single cooperator we get a single value of the structure coefficient, but  fixation probabilities  vary over initial configurations as shown for the Frucht and for the Tietze graph~\cite{mcavoy15}.

All these considerations show that calculating fixation probabilities and fixation times for multiple mutant configurations is not only computationally expensive, but also has a huge number of possible setups, for instance, which one of the considerable number of graphs to analyze, or where to place cooperators on the evolutionary graph and how many. There are various experimental parameters to be taken into account, which   might be why so far systematically conducted  numerical studies are sparse. 
In this sense, another contribution of this paper might be seen in pointing at settings for numerical experiments calculating fixation probabilities and fixation times. The results given in this paper show that among all the regular interaction graphs with $N\leq 14$ players and $3 \leq k \leq N-3$ coplayers, there is a comparably small number of graphs (as given in Tab. \ref{tab:graphs1}) which favor cooperation more than others. It may be interesting to see if these graphs also stand out in terms of fixation probability and fixation time  as compared to a graph randomly drawn from the other ones.

%%%%%%%%%%%%%%%%%%%%%%%%%%%%%%%%%%%%%%%%%%
\section*{Acknowledgments:} I wish to thank Markus Meringer for making available the \texttt{genreg} software~\cite{mer99} used for generating the regular graphs according to Tab. \ref{tab:graphs} and for helpful discussions.

\section*{Appendix A Configurations, regular graphs and structure coefficients}

The co--evolutionary games we consider here have $N$ players $\mathcal{I}= \{ \mathcal{I}_i \}$, $i=1,2,\ldots,N$, that each uses either of two strategies $\pi_i \in \{C,D \}$,  which we may interpret as cooperating or defecting. Each player $\mathcal{I}_i$, which interacts with a coplayer   $\mathcal{I}_j$, receives payoff according to the $2 \times 2$ payoff matrix       
\begin{equation} 
\bordermatrix{\: {}_i \: \backslash \: {}^j & C & D \cr
                  C & a & b \cr
                  D & c & d \cr}. \label{eq:payoff}
\end{equation} 
Which player interacts with whom is described by the interaction graph  $\mathcal{G}=(V,E)$, where the vertices $v_i \in V$ represent the players and the edges $e_{ij} \in E$ indicate that the players $\mathcal{I}_i$ and $\mathcal{I}_j$ interact as mutual coplayers~\cite{lieb05,ohts07,rich17}. Which strategy is used by which player at a given point of time is specified by a configuration $\pi=(\pi_1,\pi_2,\ldots,\pi_N)$ with  $\pi_i \in \{C,D \}$. If we represent the two strategies by a binary code  $\{C,D \} \rightarrow \{1,0 \}$, a configuration appears as a binary string the Hamming weight of which denotes the number of cooperators $c(\pi)$. For games with $N$ players, there are $2^N$ configurations with 2 configurations ($\pi=(00\ldots 0)$ and $\pi=(11\ldots1)$) absorbing.  Players may update their strategies in an updating process, for instance death--birth (DB) or birth--death (BD) updating~\cite{allen14,patt15}. Recently, it was shown by Chen et al.~\cite{chen16} that strategy $\pi_i=1=C$ is favored over $\pi_i=0=D$  if \begin{equation} \sigma(\pi,\mathcal{G}) (a-d)> (c-b). \label{eq:cond} \end{equation}
This results applies  to weak selection and $2 \times 2$ games with $N$ players, payoff matrix (\ref{eq:payoff}), any configuration $\pi$ of cooperators and defectors and for any interaction network modeled by a simple, connected, $k$--regular graph.

The quantity $\sigma(\pi,\mathcal{G})$ in Eq. (\ref{eq:cond}) is the structure coefficient of the configuration $\pi$ and the graph $\mathcal{G}$. It may not have the same value for different arrangements of cooperators and defectors described by the configuration $\pi$ and also for  different interaction networks modeled by a regular graph $\mathcal{G}$.  In particular, it was shown that 
for weak selection and the graph $\mathcal{G}$ describing  interaction as well as replacement graph,  the structure coefficient $\sigma(\pi,\mathcal{G})$ can be calculated with time complexity  $\mathcal{O}(k^2N)$ for  DB and BD updating~\cite{chen16}. For DB updating there is  \begin{equation} 
\sigma(\pi,\mathcal{G})=\frac{N\left(1+1/k \right) \overline{\omega^1} \cdot \overline{\omega^0}-2\overline{\omega^{10}}-\overline{\omega^1 \omega^0} }{N\left(1-1/k \right) \overline{\omega^1} \cdot \overline{\omega^0}+\overline{\omega^1 \omega^0}},  \label{eq:sigma}
\end{equation} with 4 local frequencies  ($\overline{\omega^1}$, $\overline{\omega^0}$, $\overline{\omega^{10}}$ and $\overline{\omega^1 \omega^0}$), which depend on $\pi$ and $\mathcal{G}$, see~\cite{chen16,rich19a,rich19b} for a probabilistic interpretation of these  frequencies. Our focus here is on DB updating as it has been shown that BD updating never favors cooperation~\cite{chen16}.

\begin{table}
\caption{The numbers $\mathcal{L}_k(N)$ of simple connected $k$--regular graphs on  $N$ vertices,~\cite{mer99}, which corresponds to the number of regular interaction graphs with $N$ players and $k$  coplayers for $6 \leq N \leq 14$ and $3 \leq k \leq N-1$. Note that there is more than one graph,  $\mathcal{L}_k(N)>1$, only for $k \leq N-3$.   }
\label{tab:graphs}
\center
\begin{tabular}{|c|c|c|c|c|c|c|c|c|c|c|}

\hline
 $\: {}_k \: \backslash \: {}^N$ & 6 & 7 & 8 & 9& 10 & 11 & 12 & 13 & 14 \\
 \hline
 3 & 2 & 0 & 5 &0 & 19 & 0 & 85 & 0 & 509 \\
 \hline
 4 & 1& 2 & 6 & 16 &59 & 265 & 1.544 & 10.778 & 88.168 \\
 \hline
 5 & 1 & 0 & 3 & 0 & 60 & 0 & 7.848 & 0 & 3.459.383 \\
 \hline
 6 & 0 & 1 & 1 & 4 & 21 & 266 & 7.849 & 367.860 & 21.609.300 \\
 \hline
 7 &0&0&1&0 & 5 & 0 & 1.547 & 0 & 21.609.301 \\
 \hline
 8 &0&0&0&1 & 1 & 6 & 94 & 10.786	& 3.459.386 \\
 \hline 
 9 &0&0&0&0& 1 & 0 & 9 & 0 & 88.193 \\
 \hline
 10 &0&0&0&0 & 0 & 1 & 1 & 10 & 540 \\ 
 \hline
 11 &0&0&0&0& 0 & 0& 1&0&13 \\
 \hline
 12 &0&0&0&0& 0 & 0& 0&1&1 \\
 \hline
 13 &0&0&0&0& 0 & 0& 0&0&1 \\ 
\hline

\end{tabular}
\end{table}

\section*{Appendix B Isomorphic graphs, isomorphic configurations and cycle counts}
The structure coefficient $\sigma(\pi,\mathcal{G})$, as for instance defined for DB updating by Eq. (\ref{eq:sigma}), may vary over configurations $\pi$ and graphs $\mathcal{G}$. This suggests the question of upper and lower bounds of $\sigma(\pi,\mathcal{G})$. For a rather low number of players it appears feasible to check all  $\sigma(\pi,\mathcal{G})$, as demonstrated in the paper for $N \leq 14$ and all regular graphs with up to 14 vertices. For a $2 \times 2$ game with $N$ players, there are $2^N-2$ non--absorbing configurations $\pi$. These configurations can be grouped according to the number of cooperators $c(\pi)$, $2 \leq c(\pi) \leq N-2$. The number of simple, connected regular graphs is known for small numbers of vertices, e.g.~\cite{mer99}, see Tab. \ref{tab:graphs}. Note that these numbers apply to graphs that are all not isomorphic with each isomorphism class being represented by exactly one graph.    In other words, Tab. \ref{tab:graphs} also gives the number of isomorphism classes for all $6 \leq N \leq 14$ and $3 \leq k \leq N-1$. Isomorphism refers to the property that two graphs are structurally alike and merely differ in how the vertices and edges are named.  More precisely,  two graphs are isomorphic if there is a bijective mapping $\theta$ between their vertices which preserves adjacency~\cite{bon08}, pp. 12--14. 

\begin{figure}[t]
\centering
\includegraphics[trim = 17mm 120mm 138mm 120mm,clip, width=3.3cm, height=3.5cm]{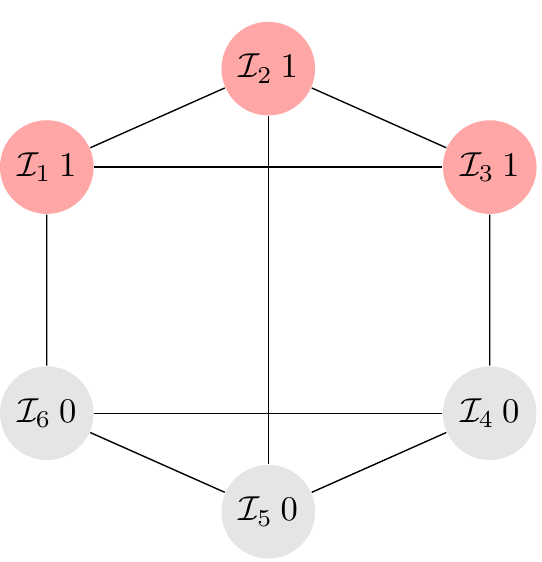} 
\includegraphics[trim = 17mm 120mm 138mm 120mm,clip, width=3.3cm, height=3.5cm]{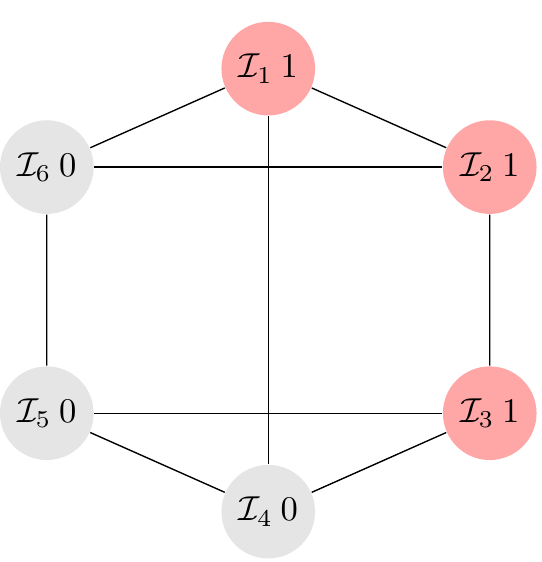} 
\includegraphics[trim = 17mm 120mm 138mm 120mm,clip, width=3.3cm, height=3.5cm]{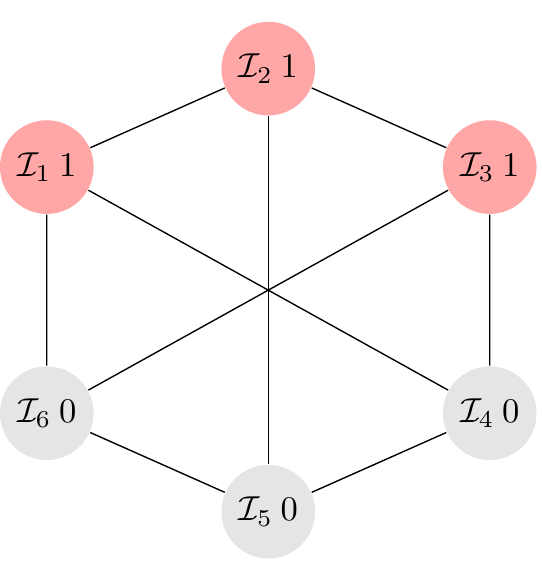}

 (a) \hspace{2.7cm} (b) $ \hspace{2.7cm}  (c) $

\caption{\small{The $\mathcal{L}_3(6)=2$ interaction graphs with $N=6$ players, each with $k=3$ coplayers. All are vertex--transitive and (c) is even symmetric (edge--transitive). The graph in (a)  has a maximal structure coefficient $\sigma_{max}=1.1818$, which is obtained for two configurations with $c(\pi)=3$ cooperators: $\pi=(111000)$ (as shown in (a)) and $\pi=(000111)$. For the isomorphic graph in (b), we get $\sigma_{max}=1.1818$ for the isomorphic configurations  $\pi=(110001)$ and $\pi=(001110)$. The graph in (c) has the same structure coefficient $\sigma=1.0000$ for all configurations. Regarding the count of cycles with length $\ell$, we see that the graphs in (a) and (b) have $\mathcal{C}_{\ell_1}(6,3)=(2,3,6,2)$, while for the graph in (c) there is $\mathcal{C}_{\ell_2}(6,3)=(0,9,0,6)$.   }}
\label{fig:graph_6_3}
\end{figure}

Consider, for example, the $\mathcal{L}_3(6)=2$ interaction graphs with $N=6$ players, each with $k=3$ coplayers, see Fig. \ref{fig:graph_6_3}. For the graph in Fig. \ref{fig:graph_6_3}a
we get the maximal structure coefficient $\sigma_{max}=1.1818$ for 2 configurations, $\pi=(111000)$ as shown in Fig. \ref{fig:graph_6_3}a and $\pi=(000111)$. By the isomorphism $\theta=\left(\begin{smallmatrix}v_1 &v_2&v_3&v_4&v_5&v_6 \\  1 &2&3&4&5&6 \\ 6 & 1 &2 &3&4&5 \end{smallmatrix} \right)$, we obtain an isomorphic graph as shown in Fig. \ref{fig:graph_6_3}b. For this graph, the configuration $\pi=(111000)$ has $\sigma=1.0000$, but  $\pi=(110001)$ and $\pi=(001110)$ have $\sigma_{max}=1.1818$. Note that between the configurations with  $\sigma_{max}$ the same isomorphic mapping $\theta$ applies. In other words, the structure coefficients are invariant under isomorphic mappings.  
For each pair of isomorphic graphs, there are isomorphic configurations that have the same value of the structure coefficient. For the graph in Fig. \ref{fig:graph_6_3}c, we obtain the result that the structure coefficient is constant over all configurations (except the absorbing configurations). Thus, isomorphic transformations do not alter the values of $\sigma(\pi,\mathcal{G})$.

These results apply generally to structure coefficients $\sigma(\pi,\mathcal{G})$ of regular graphs. The local frequencies in Eq. (\ref{eq:sigma}) solely depend on counting two types of paths on the interaction graph~\cite{chen16,rich19a,rich19b}. The quantities $\overline{\omega^1}$, $\overline{\omega^0}$ and $\overline{\omega^1 \omega^0}$ relate to the number of paths with length 1 that connect any vertex with adjacent vertices that hold a cooperator (or defector). The quantity  $\overline{\omega^{10}}$ relates to the number of paths with length 2 from any vertex to adjacent vertices on which the first vertex of the path holds a cooperator and the second vertex holds a defector.  As an isomorphic reshuffling of vertices preserves adjacency,  these numbers stay the same if the isomorphism acts on both the vertices and the configurations. 
Thus, suppose two graphs $\mathcal{G}_i$ and $\mathcal{G}_j$ are isomorphic with isomorphism $\theta$. Then, it follows $\sigma(\pi,\mathcal{G}_i)=\sigma(\theta(\pi),\mathcal{G}_j)$. Furthermore, the maximal structure coefficient is invariant as well, that is for isomorphic  graphs $\mathcal{G}_i$ and $\mathcal{G}_j$ there is
$\sigma_{max_i}=\underset{\pi}{\max} \: \sigma(\pi,\mathcal{G}_i)=\sigma_{max_j}=\underset{\pi}{\max} \: \sigma(\pi,\mathcal{G}_j)$.
Any regular graph belongs to one of the isomorphism classes and can be obtained by isomorphic transformations by any member of this class. Regular interaction graphs that are isomorphic have the same distribution of structure coefficients $\sigma(\pi,\mathcal{G})$ over the number of cooperators $c(\pi)$. Thus, by considering one representative of each isomorphism class, we can make statements about structure coefficients for all regular graphs.

For each graph, there is a specific count $\mathcal{C}_\ell(N,k)$ of cycles with length $\ell$, $3 \leq \ell \leq N$. There are efficient algorithms to count these cycles~\cite{gis19}. Consider again the $\mathcal{L}_3(6)=2$ graphs with $N=6$ players and $k=3$ coplayers, see Fig. \ref{fig:graph_6_3}. 
We find the graph in Fig. \ref{fig:graph_6_3}a and Fig. \ref{fig:graph_6_3}b has $\mathcal{C}_{\ell_1}(6,3)=(2,3,6,2)$ with $\ell=\{3,4,5,6\}$ (there are 2 cycles of length $\ell=3$, 3 cycles of length $\ell=4$, 6 cycles of length $\ell=5$  and so on), while  the graph in Fig. \ref{fig:graph_6_3}c has $\mathcal{C}_{\ell_2}(6,3)=(0,9,0,6)$. It generally applies that isomorphic graphs have the same $\mathcal{C}_\ell(N,k)$. Graphs that are not isomorphic have frequently a distinct count  $\mathcal{C}_\ell(N,k)$, but there are also cases, particularly for $N$ getting larger, where 2 not isomorphic graphs have the same  count  $\mathcal{C}_\ell(N,k)$.

\section*{Appendix C Collection of $\sigma_{max}$--graphs with $N \leq 14$}
We here give a collection of selected  $\sigma_{max}$--graphs with $N \leq 14$. The graphs are shown to illustrate some graph--theoretical properties associated with prevalence of cooperation.    The single  $\sigma_{max}$--graph with $N =6$ is already shown in Fig. \ref{fig:graph_6_3}a. For $N=7$, there are $\mathcal{L}_4(7)=2$ regular graph, which both have the same maximal structure coefficients. In other words, the count of graphs equals the count of $\sigma_{max}$--graph, which is why they are not included in the collection.
\begin{figure}[t]
\centering
\includegraphics[trim = 17mm 120mm 138mm 120mm,clip, width=3.3cm, height=3.5cm]{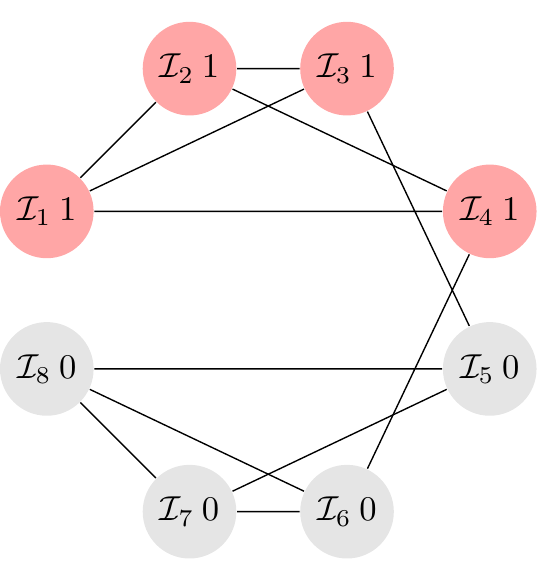} 
\includegraphics[trim = 17mm 120mm 138mm 120mm,clip, width=3.3cm, height=3.5cm]{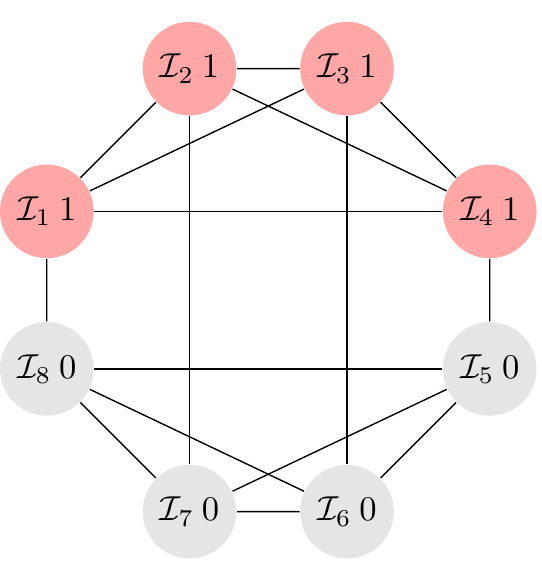} 
\includegraphics[trim = 17mm 120mm 138mm 120mm,clip, width=3.3cm, height=3.5cm]{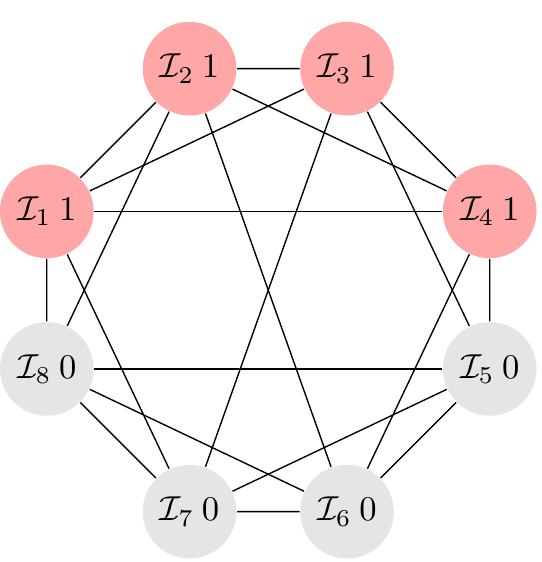} 
\includegraphics[trim = 17mm 120mm 138mm 120mm,clip, width=3.3cm, height=3.5cm]{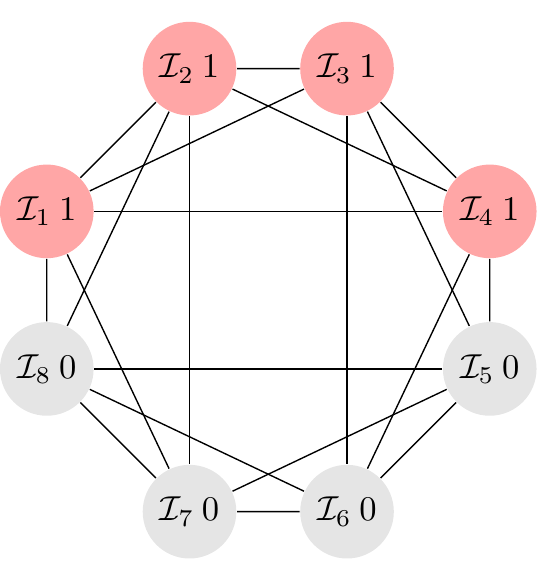}

 (a) $k=3$  \hspace{1.75cm} (b) $k=4$   \hspace{1.75cm}  (c) $k=5$  \hspace{1.75cm}  (d) $k=5$

\caption{\small{The $\sigma_{max}$--graphs for $N=8$ and $k=3,4,5$. We get $\sigma_{max}=1.6538$ for $k=3$, (a), $\sigma_{max}=1.2222$ for $k=2$, (b), and $\sigma_{max}=0.9565$ for the 2 $\sigma_{max}$--graphs with $k=5$, (c),(d), each  for the configuration $\pi=(1111 \: 0000)$. In addition, the same structure coefficient is obtained also for the configuration $\pi=(0000 \: 1111)$}, and only for (d) additionally for  $\pi=(1100 \: 0011)$ and $\pi=(0011 \: 1100)$.}
\label{fig:graph_8_x}
\end{figure}

\begin{figure}[t]
\centering
\includegraphics[trim = 17mm 120mm 132mm 120mm,clip, width=3.3cm, height=3.5cm]{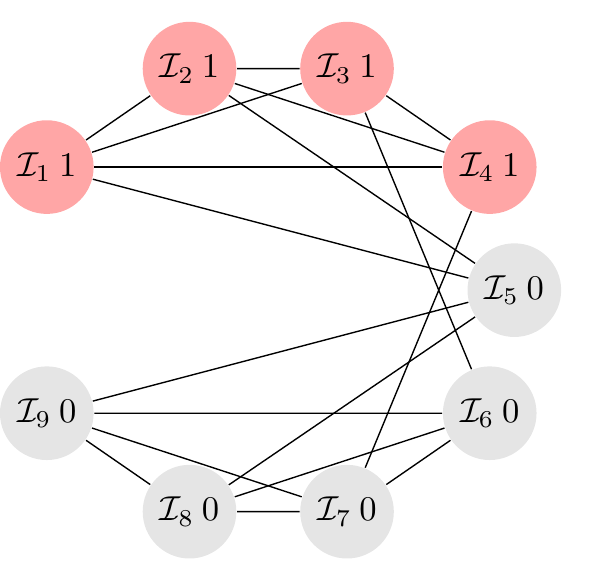} 
\includegraphics[trim = 17mm 120mm 132mm 120mm,clip, width=3.3cm, height=3.5cm]{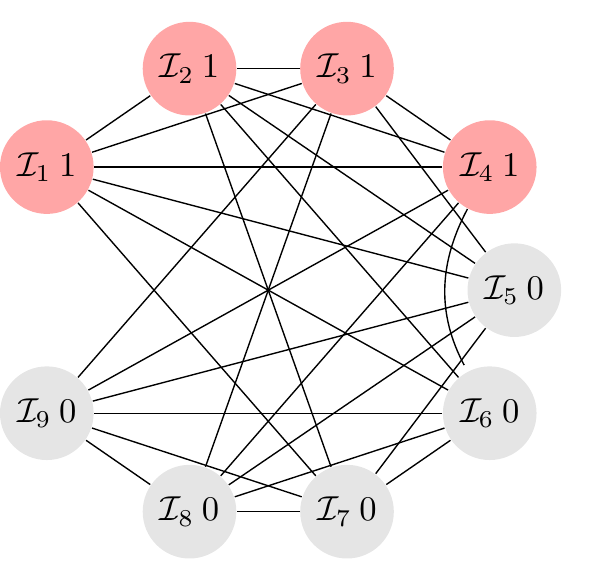} 
\includegraphics[trim = 17mm 120mm 132mm 120mm,clip, width=3.3cm, height=3.5cm]{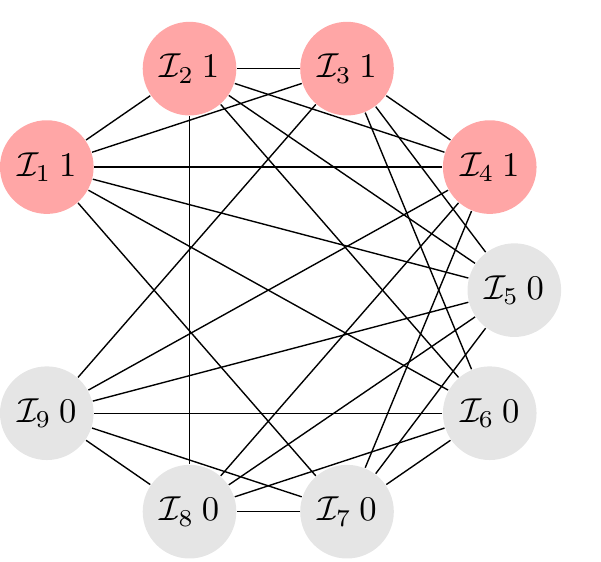} 
\includegraphics[trim = 17mm 120mm 132mm 120mm,clip, width=3.3cm, height=3.5cm]{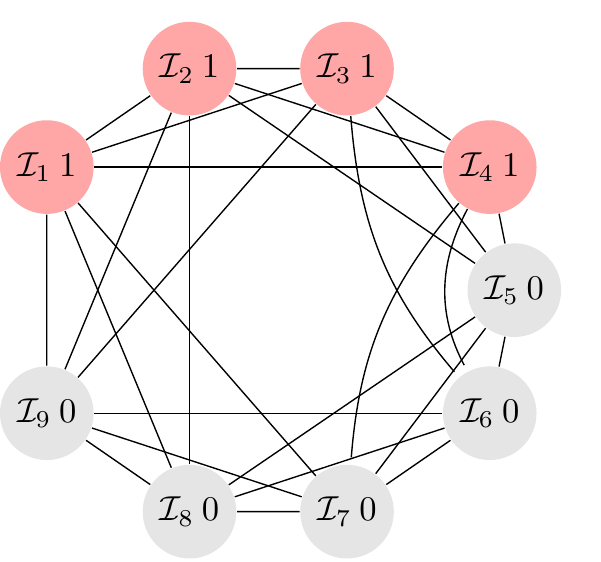}

 (a) $k=4$ \hspace{1.75cm} (b) $k=6$  \hspace{1.75cm}  (c) $k=6$  \hspace{1.75cm}  (d) $k=6$

\caption{\small{The $\sigma_{max}$--graphs for $N=9$ and $k=4,6$. We get $\sigma_{max}=1.3206$ for $k=4$ (a) and the configuration $\pi=(11110 \: 0000)$, but also for $\pi=(11111 \: 0000)$, $\pi=(00000 \: 1111)$ and $\pi=(00001 \: 1111)$. For $k=6$, there are 3 $\sigma_{max}$--graphs, (b),(c),(d), each with $\sigma_{max}=0.9115$ for the configuration $\pi=(11110 \: 0000)$. There are several more configurations that have the same $\sigma_{max}$ due to the symmetry properties of these 3 graphs.  } }
\label{fig:graph_9_x}
\end{figure}

\begin{figure}[t]
\centering
\includegraphics[trim = 17mm 120mm 132mm 120mm,clip, width=3.3cm, height=3.5cm]{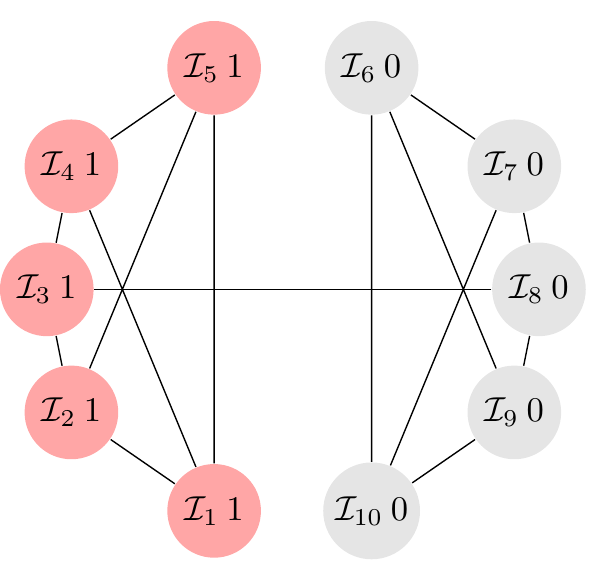} 
\includegraphics[trim = 17mm 120mm 132mm 120mm,clip, width=3.3cm, height=3.5cm]{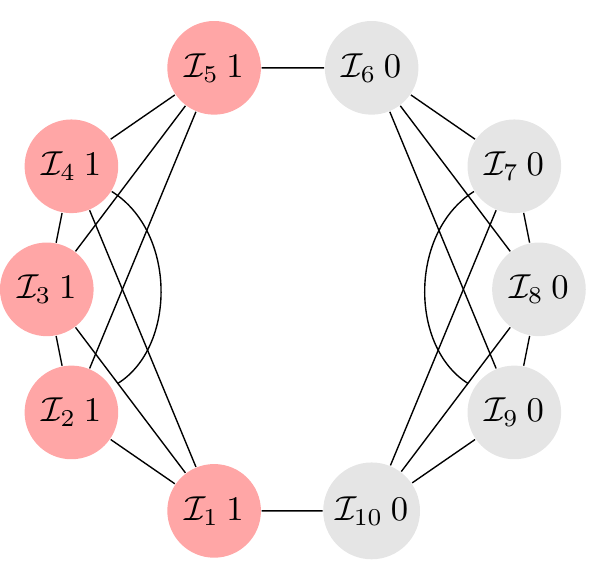} 
\includegraphics[trim = 17mm 120mm 132mm 120mm,clip, width=3.3cm, height=3.5cm]{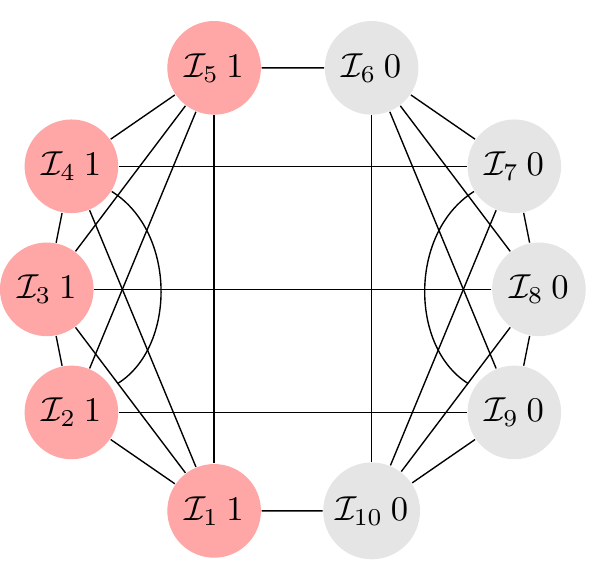}

 (a) $k=3$ \hspace{1.85cm} (b) $k=4$   \hspace{1.85cm}  (c)  $k=5$

\includegraphics[trim = 17mm 120mm 132mm 120mm,clip, width=3.3cm, height=3.5cm]{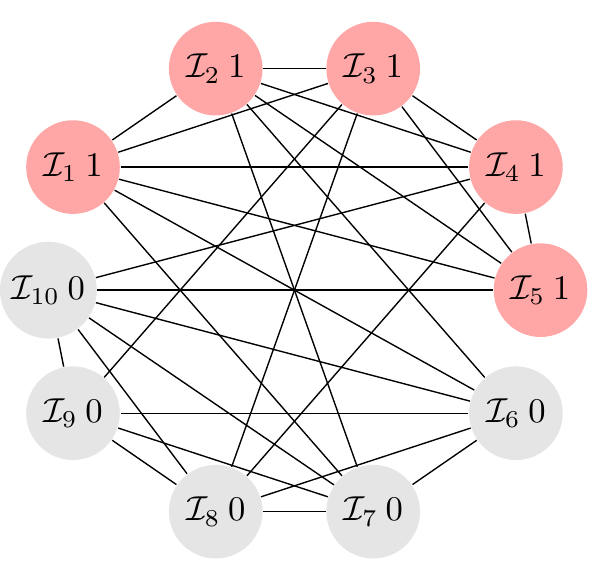} 
\includegraphics[trim = 17mm 120mm 132mm 120mm,clip, width=3.3cm, height=3.5cm]{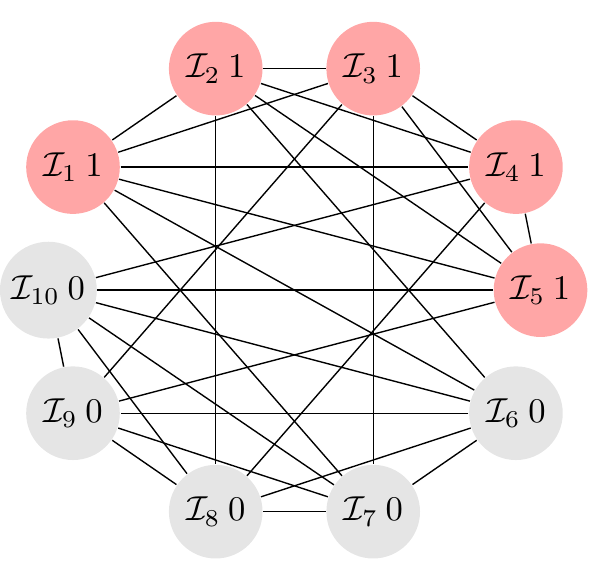} 
\includegraphics[trim = 17mm 120mm 132mm 120mm,clip, width=3.3cm, height=3.5cm]{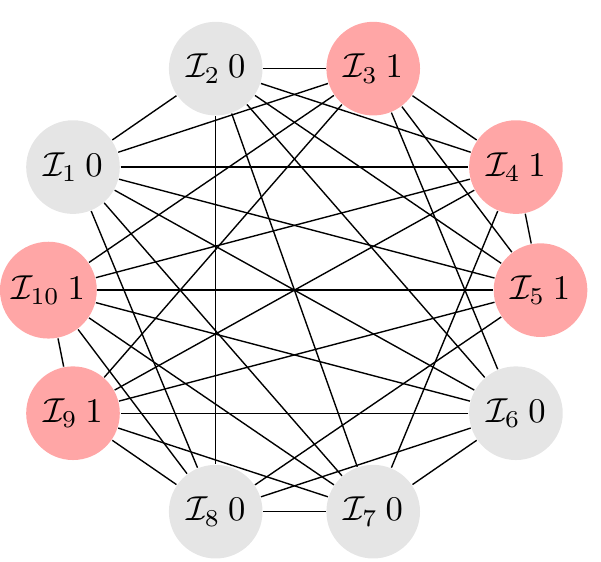} 
\includegraphics[trim = 17mm 120mm 132mm 120mm,clip, width=3.3cm, height=3.5cm]{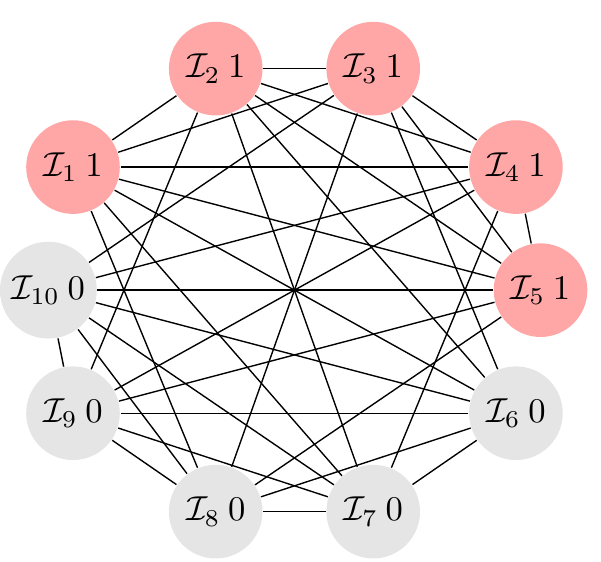}

 (d) $k=6$  \hspace{1.75cm} (e) $k=6$  \hspace{1.75cm}  (f)  $k=7$ \hspace{1.75cm}  (g) $k=7$

\caption{\small{The $\sigma_{max}$--graphs for $N=10$ and $k=3,4,\ldots,7$. We get $\sigma_{max}=1.8831$ for $k=3$ (a), $\sigma_{max}=1.5128$ for $k=4$ (b), $\sigma_{max}=1.2222$ for $k=5$ (c), $\sigma_{max}=1.0241$ for $k=6$ (d),(e)  and $\sigma_{max}=0.9145$ for $k=7$ (f),(g), all for
the configuration $\pi=(11111 \: 00000)$, and also for $\pi=(00000 \: 11111)$. For one graph with $k=7$,(f) the maximal structure coefficient is also obtained for 2 more configurations. } }
\label{fig:graph_10_x}
\end{figure}

\begin{figure}[t]
\centering
\includegraphics[trim = 17mm 120mm 138mm 120mm,clip, width=3.3cm, height=3.5cm]{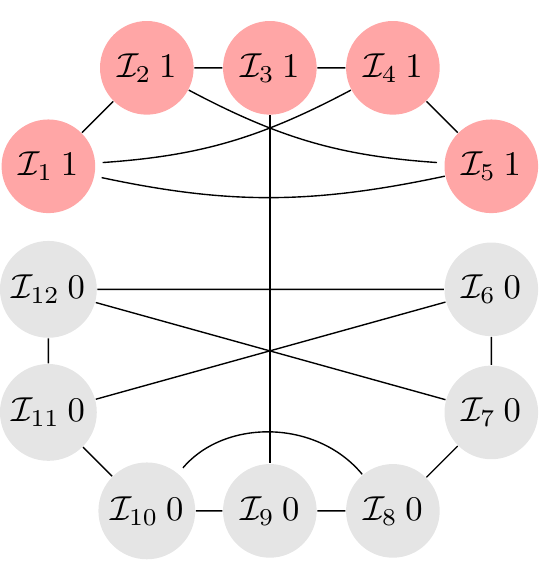} 
\includegraphics[trim = 17mm 120mm 138mm 120mm,clip, width=3.3cm, height=3.5cm]{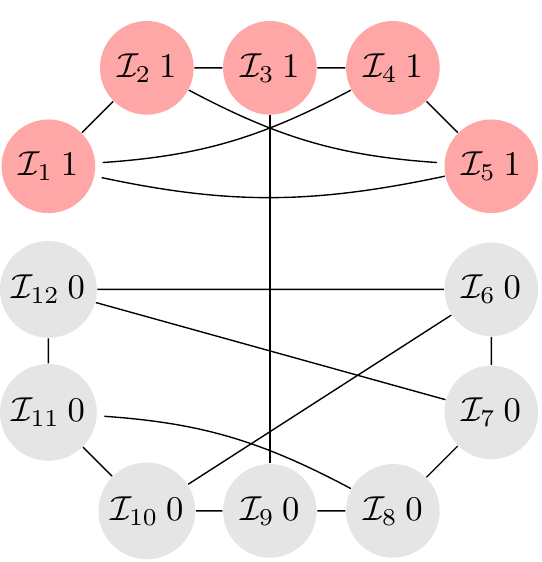} 
\includegraphics[trim = 17mm 120mm 138mm 120mm,clip, width=3.3cm, height=3.5cm]{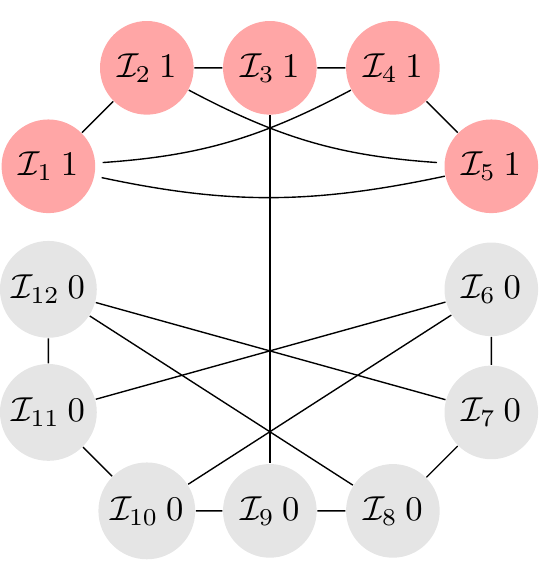} 
\includegraphics[trim = 17mm 120mm 138mm 120mm,clip, width=3.3cm, height=3.5cm]{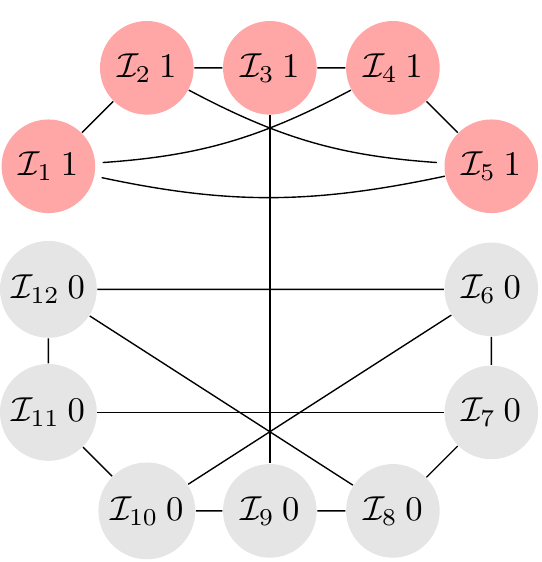}

 (a)  \hspace{2.8cm} (b)   \hspace{3cm}  (c)   \hspace{2.8cm}  (d)

\caption{\small{The $\sigma_{max}$--graphs for $N=12$ and $k=3$, each with $\sigma_{max}=1.9159$ for the configuration $\pi=(1111 \: 1000 \: 0000)$ (and also for $\pi=(0000 \: 0111 \: 1111)$) . }}
\label{fig:graph_12_3}
\end{figure}

\begin{figure}[t]
\centering
\includegraphics[trim = 17mm 120mm 138mm 120mm,clip, width=3.3cm, height=3.5cm]{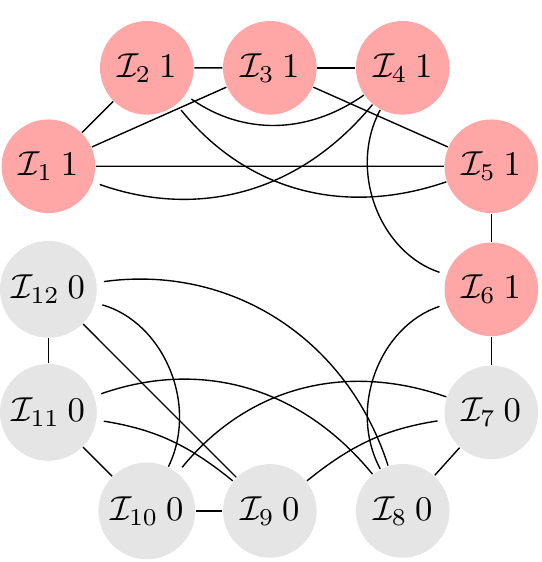} 
\includegraphics[trim = 17mm 120mm 138mm 120mm,clip, width=3.3cm, height=3.5cm]{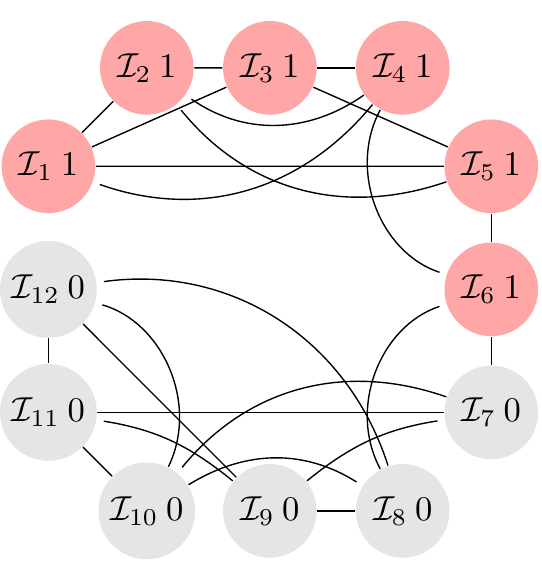} 
\includegraphics[trim = 17mm 120mm 138mm 120mm,clip, width=3.3cm, height=3.5cm]{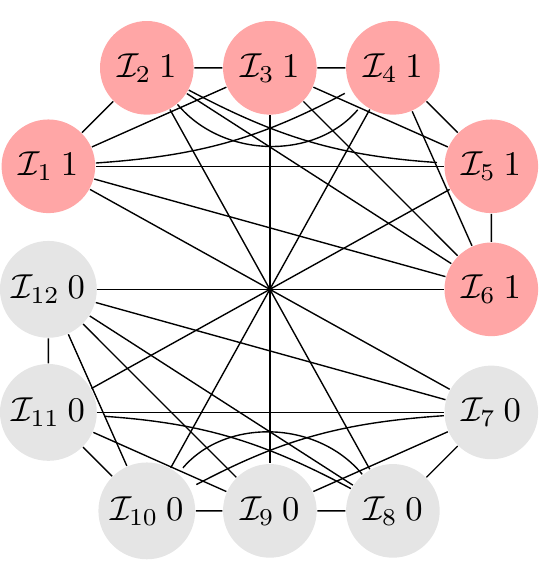} 

 (a) $k=4$ \hspace{1.85cm} (b)  $k=4$ \hspace{1.85cm} (c)   $k=6$

\caption{\small{The $\sigma_{max}$--graphs for $N=12$ and $k=4,6$. We have $\sigma_{max}=1.5701$ for $k=4$ (a),(b) and  $\sigma_{max}=1.2105$ for $k=6$ (c), each for the configuration $\pi=(1111 \: 1100 \: 0000)$ (and also for $\pi=(0000 \: 0011 \: 1111)$). }}
\label{fig:graph_12_4}
\end{figure}

\begin{figure}[t]
\centering
\includegraphics[trim = 15mm 115mm 128mm 100mm,clip, width=3.8cm, height=4.5cm]{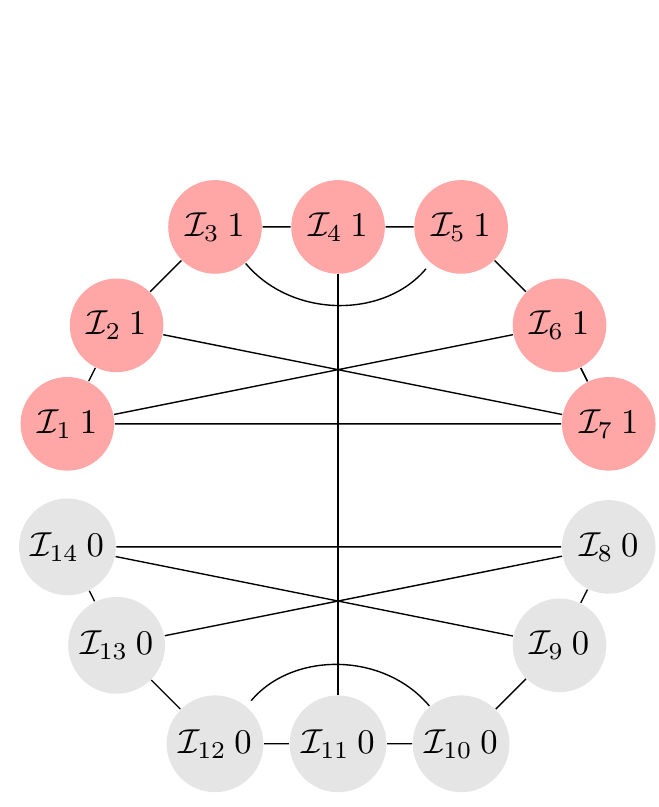} 
\includegraphics[trim = 15mm 115mm 128mm 100mm,clip, width=3.8cm, height=4.5cm]{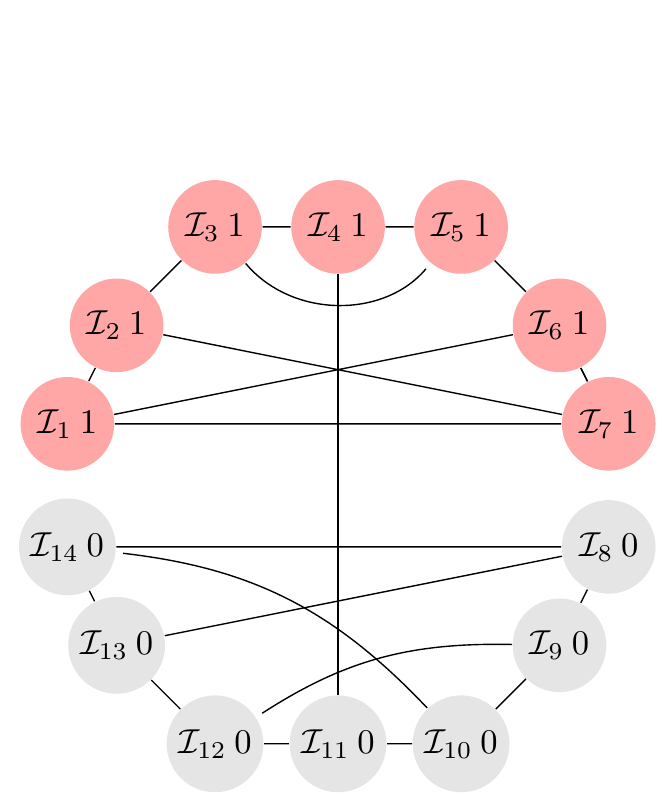} 
\includegraphics[trim = 15mm 115mm 128mm 100mm,clip, width=3.8cm, height=4.5cm]{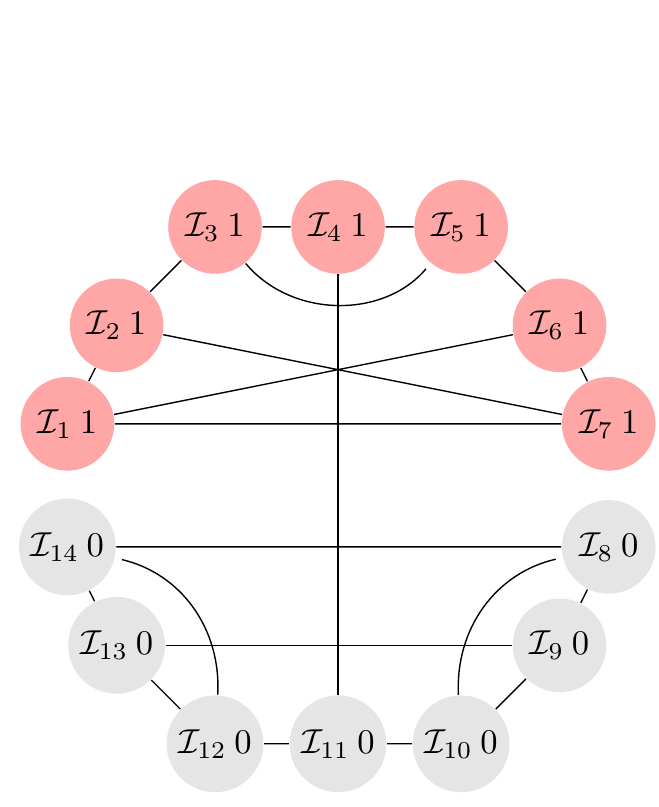} 
\includegraphics[trim = 15mm 115mm 128mm 100mm,clip, width=3.8cm, height=4.5cm]{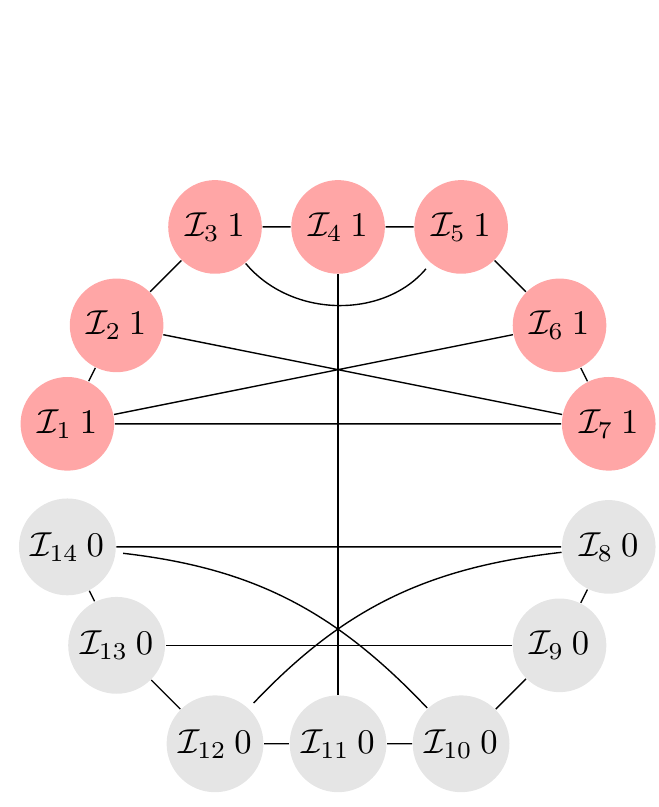} 

 (a)  \hspace{2.8cm} (b)  \hspace{2.8cm}  (c)  \hspace{2.8cm} (d)   

\includegraphics[trim = 15mm 115mm 128mm 100mm,clip, width=3.8cm, height=4.5cm]{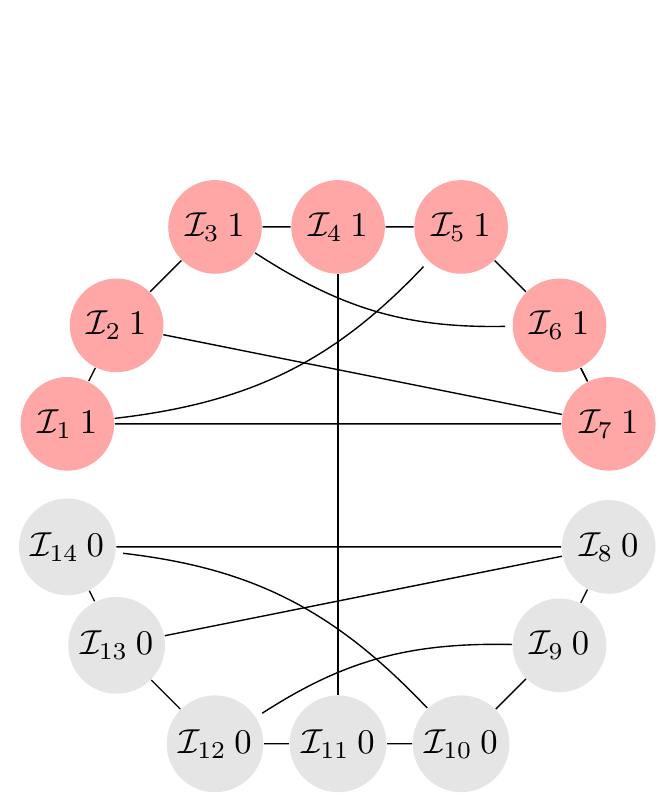} 
\includegraphics[trim = 15mm 115mm 128mm 100mm,clip, width=3.8cm, height=4.5cm]{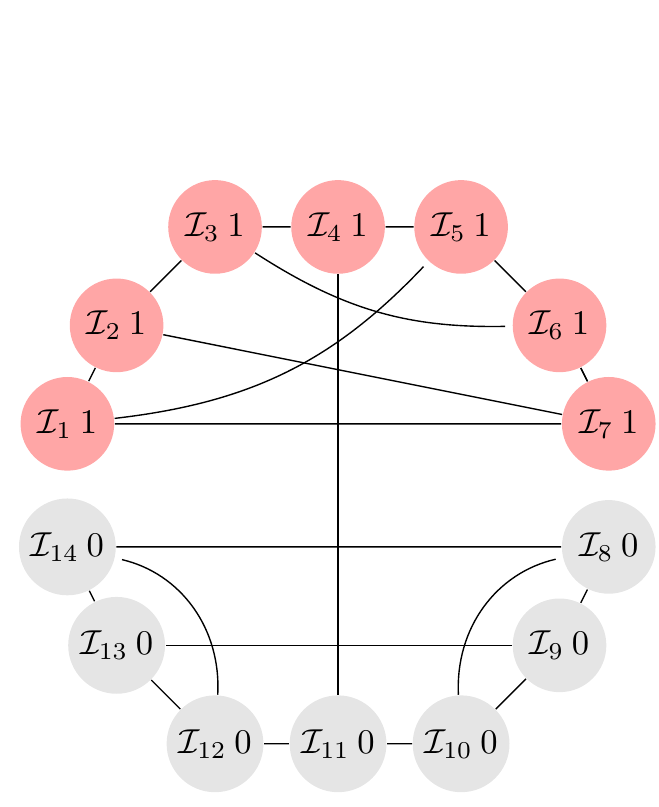} 
\includegraphics[trim = 15mm 115mm 128mm 100mm,clip, width=3.8cm, height=4.5cm]{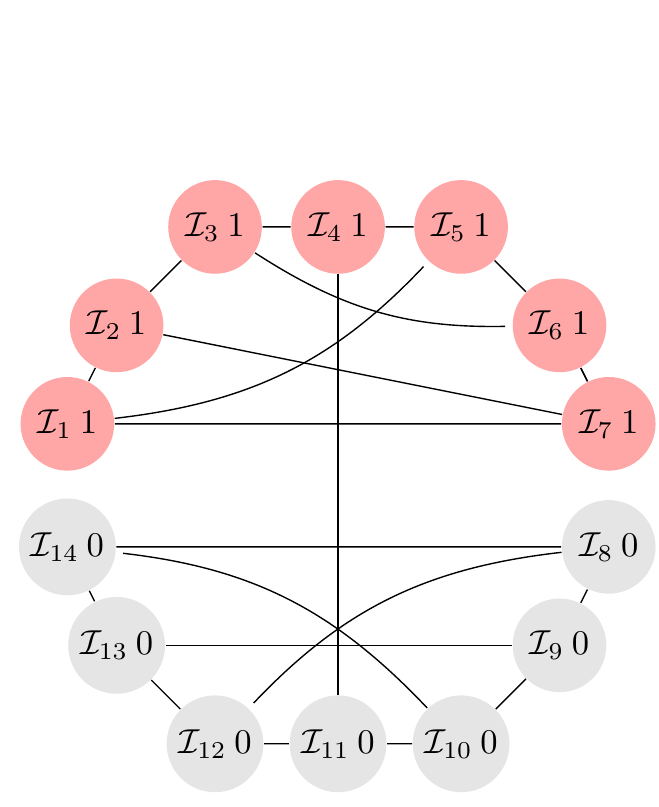} 
\includegraphics[trim = 15mm 115mm 128mm 100mm,clip, width=3.8cm, height=4.5cm]{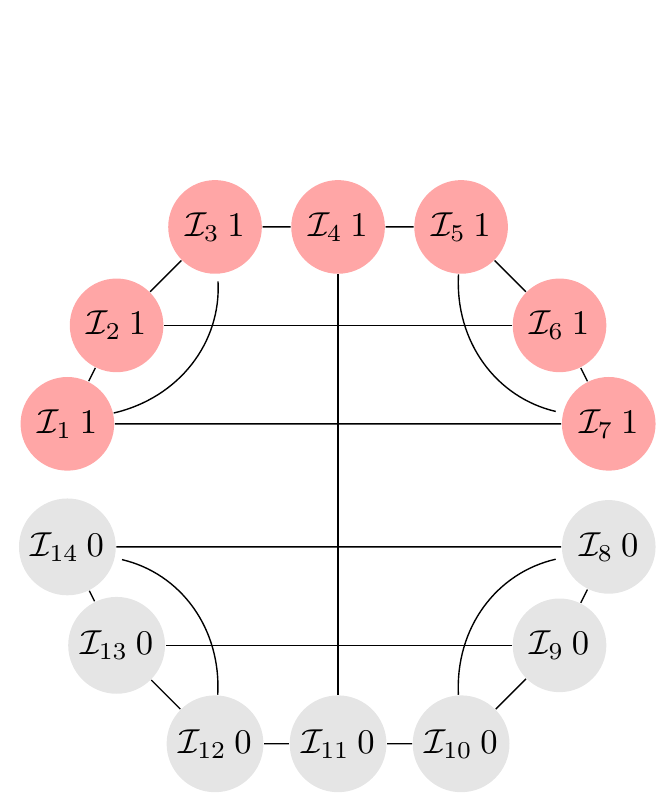} 

 (e)  \hspace{2.8cm} (f) \hspace{2.8cm}   (g)  \hspace{2.8cm} (h)   
 
\caption{\small{The $\sigma_{max}$--graphs for $N=14$ and $k=3$, each with $\sigma_{max}=1.9396$ for the configuration $\pi=(1111 \: 1110 \: 0000 \: 00)$ (and also for  $\pi=(0000 \: 0001 \: 1111 \: 11)$) . Only 8 out of $\#_{sigma_{max}}=10$ $\sigma_{max}$--graphs according to Tab. \ref{tab:graphs1} are depicted. The remaining 2 graphs arise from the blocks depicted in the figures. If we call the upper half of the graphs in Fig. \ref{fig:graph_14_3} (a)--(d) the A--block, then the lower half of these graphs consists of blocks A, B, C and D. The blocks are joined by the edge connecting $\mathcal{I}_4$ and  $\mathcal{I}_{11}$. The graphs in \ref{fig:graph_14_3} (e)--(f) also are block--like with blocks B--B, B--C, B--D and C--C. The 2 remaining graphs are formed by connecting the blocks C--D and D--D. }}
\label{fig:graph_14_3}
\end{figure}

Figs. \ref{fig:graph_8_x}--\ref{fig:graph_10_x} shown all $\sigma_{max}$--graphs for $N=8,9,10$ and $3 \leq k \leq N-3$ together with $\sigma_{max}$ and the  associated configurations. For $N=12$ and $N=14$, only some examples of $\sigma_{max}$--graphs are given in Figs. \ref{fig:graph_12_3}--\ref{fig:graph_14_3}
due to brevity.
A full list of all $\sigma_{max}$--graphs for $11 \leq N \leq 14$ and $3 \leq k \leq N-3$ is made available here~\cite{rich20}.
It is particularly noticeable that the $\sigma_{max}$--graphs   
are structured to have blocks with clusters of mutants. For instance, we see such a block with $(\mathcal{I}_1,\mathcal{I}_2,\mathcal{I}_3,\mathcal{I}_4)$ for the graph with $N=8$ and $k=3$ in Fig. \ref{fig:graph_8_x}a and for $N=9$ and $k=4$ in  Fig. \ref{fig:graph_9_x}a, or for $N=10$ and $k=3$, Fig. \ref{fig:graph_10_x}a and for the cubic graphs ($k=3$) with $N=12$ and $N=14$ as well, see  Figs. \ref{fig:graph_12_3} and \ref{fig:graph_14_3}. The $\sigma_{max}$--graphs with larger degree ($=$ coplayers) still somewhat retains such a ``blockish'' appearance (for instance $(\mathcal{I}_1,\mathcal{I}_2,\mathcal{I}_3,\mathcal{I}_4,\mathcal{I}_5)$ in Fig. \ref{fig:graph_10_x}c) but to a far lesser degree.  In addition,  $\sigma_{max}$--graphs with larger degree are frequently vertex--transitive (for instance Figs. \ref{fig:graph_9_x}d,  \ref{fig:graph_10_x}e and \ref{fig:graph_10_x}g) which is not the case for cubic ($k=3$) and quartic ($k=4$)  $\sigma_{max}$--graphs with $N \leq 14$, with the exception of $N=6$ and $k=3$, see Fig. \ref{fig:graph_6_3}a.  Furthermore, it can be observed that
the blocks are occupied by clusters of cooperators which are frequently connected by cut vertices and/or hinge vertices. 
For instance, for $N=12$ and $k=3$, the vertices occupied by the players $\mathcal{I}_3$ and $\mathcal{I}_9$, see Fig. \ref{fig:graph_12_3}, are cut vertices, while for $N=10$ and $k=4$,  see Fig. \ref{fig:graph_10_x}b, the vertices occupied by the players $\mathcal{I}_5$ and $\mathcal{I}_6$ are  hinge vertices as their removal would make the distance between  $\mathcal{I}_4$ and  $\mathcal{I}_7$ longer.  As discussed above,
the clusters can be seen as to serve as a mutant family that invades the remaining graph. As vertices with players of opposing strategies are connected by cut and/or hinge vertices there is only a small number of (or even just a single) migration path for the cooperators and/or defectors.

\end{document}